\documentclass[12pt,a4paper]{article}
\usepackage{amsmath}
\usepackage{amssymb}
\usepackage[top=2cm, bottom=2cm, left=2cm, right=2cm]{geometry}
\usepackage{indentfirst}
\usepackage{hyperref}
\usepackage[numbers,sort&compress]{natbib}
\usepackage{color}
\usepackage{graphicx}

\allowdisplaybreaks[4]

\begin{document}

\renewcommand{\thefootnote}{\fnsymbol{footnote}}

\begin{titlepage}

\vspace{10mm}
\begin{center}
{\Large\bf All-Loop Renormalizable Wess-Zumino Model on Bosonic-Fermionic Noncommutative Superspace}
\vspace{16mm}

{\large Yan-Gang Miao${}^{1,2,}$\footnote{E-mail address: miaoyg@nankai.edu.cn} and Xu-Dong Wang${}^{1,}$\footnote{E-mail address: c\_d\_wang@mail.nankai.edu.cn}

\vspace{6mm}
${}^{1}${\normalsize \em School of Physics, Nankai University, Tianjin 300071, China}

\vspace{3mm}
${}^{2}${\normalsize \em Bethe Center for Theoretical Physics and Institute of Physics, University of Bonn, \\
Nussallee 12, D-53115 Bonn, Germany}}

\end{center}

\vspace{10mm}
\centerline{{\bf{Abstract}}}
\vspace{6mm}

We generalize the ordinary Wess-Zumino model to the Bosonic-Fermionic noncommutative (BFNC) superspace and study its renormalization. In our previous work that can be regarded as the key foundation of the present paper, we have proved that the BFNC Wess-Zumino model with the real  mass and  interacting constant is one-loop renormalizable up to the second order of  BFNC parameters. Based on the result obtained, in the present paper we modify the one-loop renormalizable  BFNC Wess-Zumino model by generalizing the mass and  interacting constant to complex numbers, introduce the $U(1)_{\rm R}$ $R$-symmetry and $U(1)_{\Phi}$ flavor symmetry in the modified model, analyze possible divergent operators in the effective action of the modified model by using the dimensional analysis method, and further give a new BFNC Wess-Zumino model that  is renormalizable at all loops still up to the second order of  BFNC parameters by imposing symmetries rather than doing a direct perturbative investigation.

\vskip 20pt
\noindent
{\bf PACS Number(s)}:  12.60.Jv, 11.30.Pb, 11.10.Nx, 11.10.Gh


\end{titlepage}

\newpage
\renewcommand{\thefootnote}{\arabic{footnote}}
\setcounter{footnote}{0}
\setcounter{page}{2}

\section{Introduction}

By deforming the ordinary spacetime and superspace to a noncommutative (NC) spacetime and non-anticommutative (NAC) superspace, respectively, and then constructing physical models on the NC spacetime and NAC superspace, one can acquire~\cite{Ferrara:2000mm, Klemm:2001yu, Ferrara:2003xy} a deeper understanding of quantum field theory. 
One of the most interesting properties is the renormalization of quantum field theory on the NC spacetime and NAC superspace. 
For instance, see ref.~\cite{Seiberg:2003yz}, on the one hand the simplest NAC superspace was given and the possibility to construct the Bosonic-Fermionic noncommutative (BFNC) superspace\footnote{The Bosonic coordinates no longer commute with the Fermionic coordinates on such a superspace.} 
was pointed out, and on the other hand  the Wess-Zumino model and the super Yang-Mills model were generalized to the NAC superspace and the effect of NAC deformation on the physical models was investigated. 
In addition, in the later researches~\cite{Britto:2003aj, Grisaru:2003fd, Britto:2003kg, Romagnoni:2003xt} the NAC Wess-Zumino model was proved to be renormalizable at one loop and then at all loops. Here we note that the NC spacetime and NAC superspace are of string theory basic. In fact, 
from the point of view of string theory~\cite{Seiberg:1999vs,Ooguri:2003qp,Ooguri:2003tt,Berkovits:2003kj}, the NC spacetime and NAC superspace can be obtained under certain conditions and furthermore the BFNC superspace~\cite{de Boer:2003dn} can be predicted.

Compared with a lot of studies on the models based on the NAC superspace, the investigations related to the BFNC superspace are few. 
In our recent work~\cite{Miao:2013a} we have constructed a one-loop renormalizable Wess-Zumino model on the BFNC superspace.
Motivated by our previous result, we go ahead in the present paper. That is, we modify the one-loop renormalizable BFNC Wess-Zumino model  by introducing a complex mass and a complex interacting, find out its effective action, and propose a new BFNC Wess-Zumino model with renormalizability at all loops.

The key point to construct an all-loop renormalizable Wess-Zumino action on the BFNC superspace is the introduction of global $U(1)$ symmetries, which has been realized~\cite{Britto:2003aj, Britto:2003kg, Romagnoni:2003xt} for the NAC Wess-Zumino action. Moreover, the prerequisite for introducing global $U(1)$ symmetries is the complexification of mass and interacting constant in our one-loop renormalizable Wess-Zumino action~\cite{Miao:2013a}, which is inspired by the realization of one- and all-loop renormalizable NAC Wess-Zumino actions. A more detailed description is given below.

The deformed Wess-Zumino action defined on the NAC superspace is invariant under an ${\cal N}=1/2$ instead of ${\cal N}=1$ supersymmetry transformation because the definition of NAC star product contains the supersymmetry generator $Q$ which does not anti-commute with the supersymmetry generator $\bar{Q}$.
In addition, besides the ${\cal N}=1/2$ supersymmetry, two global $U(1)$ symmetries are found~\cite{Britto:2003aj, Britto:2003kg, Romagnoni:2003xt} for the NAC Wess-Zumino action with a complex mass and a complex interacting constant, and they play a crucial role in the construction of one- and all-loop renormalizable NAC Wess-Zumino actions.
Inspired by the success in the NAC case together with the existence of the ${\cal N}=1/2$ supersymmetry in the BFNC case, we generalize the mass and interacting constant to complex numbers in our one-loop renormalizable BFNC Wess-Zumino action. Again considering that a global $U(1)$ symmetry is independent of supersymmetry, we then define the $U(1)_{\rm R}$ $R$-symmetry and $U(1)_{\Phi}$ flavor symmetry charges for the complex mass and interacting constant.
Besides, we need to define the symmetry charges for the BFNC parameters and operators listed in Table 1.

We emphasize that the renormalization analysis based on the global $U(1)$ symmetries leads to renormalizability at all loops because it is not related to a direct perturbative process. Moreover, the global $U(1)$ symmetries can greatly simplify this analysis of renormalization. The reason is that if an action has global $U(1)$ symmetries and an ${\cal N}=1/2$ supersymmetry, based on the background field method~\cite{Gates:1983nr}, its effective action also has these symmetries.
As a result, one can seek out the forms of divergent operators allowed by the symmetries from all possible forms in an effective action, and furthermore determine the exact form of an effective action.
We shall see that such an analysis is indeed very powerful to construct the renormalizable BFNC Wess-Zumino action we desire.

The paper is organized as follows. In the next section we give a detailed summary of the previous paper~\cite{Miao:2013a}, both for conventions, definitions and the main results in order to make the present paper self-consistent because this paper is intimately related to the previous one. Then we demonstrate our main idea and specific treatment on the investigation of renormalizability of a new BFNC Wess-Zumino model at all loops in section 3.
The following five sections thus focus on technical jobs. Based on the one-loop renormalizable Wess-Zumino action ${\cal S}_{(3)}$~\cite{Miao:2013a}, we give the modified action ${\cal S}^{\prime}_{(3)}$ which has the $U(1)_{\rm R}$ $R$-symmetry and $U(1)_{\Phi}$ flavor symmetry in section 4. Next, 
we derive the constraints that should be satisfied by possible divergent operators in the effective action of ${\cal S}^{\prime}_{(3)}$ in section 5, and further determine the allowed divergent operators by solving the constraints in section 6.
We turn to the construction of BFNC parameters in section 7. In terms of the allowed forms of divergent operators and BFNC parameters we are able to write the effective action of ${\cal S}^{\prime}_{(3)}$ in section 8.
Finally,  we present our conclusion and outlook in section 9, where a new BFNC Wess-Zumino action that is renormalizable at all loops  is determined. 

\section{Summary of the Previous Paper}

We give a detailed summary of our previous work~\cite{Miao:2013a} in order for this paper to be self-consistent.

At first, the structure of the BFNC algebra takes the form,
\begin{eqnarray}
\label{algebra_relations}
\left[y^k,\theta ^{\alpha }\right]&=&i \Lambda ^{k \alpha },
\qquad 
\left[x^k,y^l\right]=-\sigma ^k{}_{\alpha  \dot{\beta }} \Lambda ^{l \alpha } \bar{\theta }^{\dot{\beta }},
\qquad 
\left[x^k,\theta ^{\alpha }\right]=i \Lambda ^{k \alpha },
\nonumber\\ 
\left[x^k,x^l\right]&=&\sigma ^l{}_{\alpha  \dot{\beta }} \Lambda ^{k \alpha } \bar{\theta }^{\dot{\beta }}-\sigma ^k{}_{\alpha  \dot{\beta }} \Lambda ^{l \alpha } \bar{\theta }^{\dot{\beta }},
\end{eqnarray}
where the other commutators and anti-commutators that do not appear in the above are vanishing. The algebraic relations give the definition of the BFNC parameters $\Lambda ^{k \alpha }$, where $k=0,1,2,3$, and $\alpha=1,2$, which implies that there are eight independent BFNC parameter components in total, and $y^k$, $\theta ^{\alpha }$, and $\bar{\theta }^{\dot{\alpha }}$ are chiral coordinates with the relation 
$y^k\equiv x^k+i \sigma ^k{}_{\alpha  \dot{\beta }} \theta ^{\alpha } \bar{\theta }^{\dot{\beta }}$. Correspondingly, 
the BFNC star product is defined 
in terms of the tensor algebraic notation which is frequently used in quantum group theory~\cite{Majid:1996kd},
\begin{eqnarray}
\label{star_product_expansion}
{\bf F} \star {\bf G}&\equiv&\mu \left\{ \exp \left[\frac{i}{2}\Lambda ^{k \alpha }\left(\frac{\partial }{\partial y^k}\otimes \frac{\partial }{\partial \theta ^{\alpha }}-\frac{\partial }{\partial \theta ^{\alpha }}\otimes \frac{\partial }{\partial y^k}\right)\right]\triangleright({\bf F}\otimes {\bf G}) \right\}, \nonumber \\
&=&{\bf F} {\bf G}-\frac{i}{2}\Lambda ^{k \alpha }\left(\partial _{\alpha }{\bf F}\right)\left(\partial _k{\bf G}\right)+(-1)^{|{\bf F}|}\frac{i}{2}\Lambda ^{k \alpha }\left(\partial _k{\bf F}\right)\left(\partial _{\alpha }{\bf G}\right) \nonumber \\
&&+\frac{1}{8}\Lambda ^{k \alpha }\Lambda ^{l \beta }\left(\partial _k\partial _l{\bf F}\right)\left(\partial _{\alpha }\partial _{\beta }{\bf G}\right)+\frac{1}{8}\Lambda ^{k \alpha }\Lambda ^{l \beta }\left(\partial _{\alpha }\partial _{\beta }{\bf F}\right)\left(\partial _k\partial _l{\bf G}\right) \nonumber \\
&&+(-1)^{|{\bf F}|}\frac{1}{4}\Lambda ^{k \alpha }\Lambda ^{l \beta }\left(\partial _{\beta }\partial _k{\bf F}\right)\left(\partial _{\alpha }\partial _l{\bf G}\right)\nonumber \\
&&-\frac{i}{16}\Lambda ^{k \alpha }\Lambda ^{l \beta }\Lambda ^{m \zeta }\left(\partial _{\alpha }\partial _l\partial _m{\bf F}\right)\left(\partial _{\beta }\partial _{\zeta }\partial _k{\bf G}\right)\nonumber \\
&&+(-1)^{|{\bf F}|}\frac{i}{16}\Lambda ^{k \alpha }\Lambda ^{l \beta }\Lambda ^{m \zeta }\left(\partial _{\alpha }\partial _{\beta }\partial _m{\bf F}\right)\left(\partial _{\zeta }\partial _k\partial _l{\bf G}\right)\nonumber \\
&&-\frac{1}{64}\Lambda ^{k \alpha }\Lambda ^{l \beta }\Lambda ^{m \zeta }\Lambda ^{n \iota }\left(\partial _{\alpha }\partial _{\zeta }\partial _l\partial _n{\bf F}\right)\left(\partial _{\beta }\partial _{\iota }\partial _k\partial _m{\bf G}\right),
\end{eqnarray}
where 
$\mu$ denotes the change from the tensor product $\otimes$ to an ordinary product, the symbol $\triangleright$ represents an action of an operator on a function, ${\bf F}$ and ${\bf G}$ stand for any superfields, $|{\bf F}|$ is the grade of ${\bf F}$ that equals zero  for a Bosonic element and one for a Fermionic element, $\partial_k\equiv \frac{\partial }{\partial y^k}$, and $\partial _{\alpha }\equiv \frac{\partial }{\partial \theta ^{\alpha }}$.
Replacing the ordinary product by the star product in the Wess-Zumino model ${\cal S}_{\rm WZ}$, we obtain the deformed Wess-Zumino model depicted by the action ${\cal S}_{{\rm NC}}$,
\begin{eqnarray}
\label{deformed WZ action}
{\cal S}_{\rm NC}&=& \int d^4x\left\{\Phi ^+\star\Phi |_{\theta ^2\bar{\theta }^2}+\frac{m}{2}\Phi  \star \Phi |_{\theta ^2}+\frac{g}{3}\Phi  \star \Phi \star \Phi |_{\theta ^2}\right.\nonumber\\
&&\hspace{15mm}\left.+\frac{m}{2}\Phi ^+ \star \Phi ^+|_{\bar{\theta }^2}+\frac{g}{3}\Phi ^+ \star  \Phi ^+ \star \Phi ^+|_{\bar{\theta }^2}
\right\},
\end{eqnarray}
where the mass $m$ and the interacting constant $g$ are real parameters, and $\Phi$ and $\Phi ^+$ are chiral and antichiral superfields defined as $\Phi \equiv \Phi (y,\theta)$ and  $\Phi ^+ \equiv \Phi ^+(y,\theta ,\bar{\theta })$, respectively.
According to the definition of the star product (see eq.~(\ref{star_product_expansion})), it is straightforward to verify that ${\cal S}_{\rm NC}$ has the $1/2$ supersymmetry defined as
\begin{equation}
\label{supersymmetry transformation}
\delta _{\xi }\Phi\equiv\xi ^{\alpha }{\partial}_{\alpha }\Phi, \qquad \delta _{\xi }\Phi ^+\equiv\xi ^{\alpha }{\partial}_{\alpha }\Phi ^+.
\end{equation}

In the BFNC superspace we can rewrite the deformed action by using the coordinates $x^k$, $\theta ^{\alpha }$, and $\bar{\theta }^{\dot{\alpha }}$,
\begin{eqnarray}
\label{S_NC}
{\cal S}_{{\rm NC}}
&=&\int d^8z \left\{\Phi ^+\Phi -\frac{m}{8}\Phi  \left(\frac{D^2}{\square }\Phi \right)-\frac{m}{8}\Phi ^+ \left(\frac{\bar{D}^2}{\square }\Phi ^+\right)\right.\nonumber\\
&&\hspace{15mm}-\frac{g}{12}\Phi  \Phi \left(\frac{D^2}{\square }\Phi \right)-\frac{g}{12}\Phi ^+ \Phi ^+ \left(\frac{\bar{D}^2}{\square }\Phi ^+\right)\nonumber\\
&&\hspace{15mm}+\frac{1}{3072}(-g) \Lambda ^{k l} \Lambda ^{n o} \theta ^4 \left(D^2 \Phi \right) \partial _l\partial _k\left(D^2 \Phi \right) \partial _o\partial _n\left(D^2 \Phi \right)\nonumber\\
&&\hspace{15mm}+\frac{1}{32} (-g) \Lambda ^{k l} \theta ^4 \Phi  \left(D^2 \Phi \right) \partial _l\partial _k\left(D^2 \Phi \right)\nonumber\\
&&\hspace{15mm}+\frac{1}{32} (-g) \Lambda ^{k l} \theta ^4 \Phi  \partial _k\left(D^2 \Phi \right) \partial _l\left(D^2 \Phi \right)\nonumber\\
&&\hspace{15mm}+\frac{1}{6} (-g) \Lambda ^{k l} \theta ^4 \Phi ^+ \square \Phi ^+ \partial _k\partial _l\Phi ^+\nonumber\\
&&\hspace{15mm}+\frac{1}{3} (-g) \left(\sigma \Lambda \Lambda ^{k l}\right)^{n o} \theta ^4 \Phi ^+ \partial _k\partial _n\Phi ^+ \partial _l\partial _o\Phi ^+\nonumber\\
&&\hspace{15mm}+\frac{1}{6} g \eta ^{k l} \Lambda ^{n o} \theta ^4 \Phi ^+ \partial _k\partial _n\Phi ^+ \partial _l\partial _o\Phi ^+\nonumber\\
&&\hspace{15mm}+\frac{1}{16} (-g) \epsilon ^{\alpha  \beta } \Lambda ^{k l} \theta ^4 \partial _k\left(D_{\alpha } \Phi \right) \partial _l\left(D_{\beta } \Phi \right) \left(D^2 \Phi \right)\nonumber\\
&&\hspace{15mm}\left.+\frac{1}{16} (-g) \epsilon ^{\alpha  \beta } \epsilon ^{\zeta  \iota } \Lambda ^k{}_{\beta } \Lambda ^l{}_{\iota } \theta ^4 \partial _k\left(D_{\alpha } \Phi \right) \partial _l\left(D_{\zeta } \Phi \right) \left(D^2 \Phi \right)
\right\},
\end{eqnarray}
where the first two lines correspond to the ordinary Wess-Zumino action ${\cal S}_{\rm WZ}$, and the remainnings are the BFNC deformed, i.e. $\Lambda$-dependent contributions denoted as ${\cal S}_{\Lambda}$ in later related expressions. 
Some relevant notations are necessary to be mentioned, 
$D_{\alpha }$ and $\bar{D}_{\dot{\alpha }}$ are defined by
\begin{eqnarray}
\label{chiral_representation}
D_{\alpha }&\equiv&\frac{\partial }{\partial \theta ^{\alpha }}+i \sigma ^k{}_{\alpha  \dot{\beta }}\bar{\theta }^{\dot{\beta }}\frac{\partial }{\partial x^k}\qquad 
\bar{D}_{\dot{\alpha }}\equiv-\frac{\partial }{\partial \bar{\theta }^{\dot{\alpha }}}-i \sigma ^k{}_{\beta  \dot{\alpha }}\theta ^{\beta }\frac{\partial }{\partial x^k},
\end{eqnarray}
and $\partial_k\equiv \frac{\partial }{\partial x^k}$, 
 $d^8z \equiv d^4x d^2\theta d^2\bar{\theta}$, $\theta ^2\equiv \theta ^{\alpha }\theta _{\alpha }$, $\bar{\theta }^2\equiv \bar{\theta }_{\dot{\alpha }}\bar{\theta }^{\dot{\alpha }}$, $\theta ^4 \equiv \theta ^2\bar{\theta }^2$, $D^2=\epsilon ^{\alpha \beta }D_{\beta }D_{\alpha }$, and $\bar{D}^2=\epsilon ^{\dot{\alpha }\dot{\beta }}\bar{D}_{\dot{\alpha }}\bar{D}_{\dot{\beta }}$.
Moreover, the following symbols are defined and have been utilized in eq.~(\ref{S_NC}) in order to make it and later related ones concise, 
\begin{eqnarray}
\label{abbrivation}
\Lambda ^{k l}&\equiv&\epsilon ^{\alpha  \beta }\Lambda ^k{}_{\beta }\Lambda ^l{}_{\alpha },\nonumber\\
\Lambda ^2&\equiv&\eta _{k l}\Lambda ^{k l},\nonumber\\
\sigma \Lambda \Lambda &\equiv&\eta _{k n}\eta _{l o}\left(\sigma ^{k l}\right)^{\alpha  \beta } \Lambda ^n{}_{\alpha } \Lambda ^o{}_{\beta },\nonumber\\
\left(\sigma \Lambda ^{k l}\right)^{n \alpha }&\equiv&\left(\sigma ^{k l}\right)^{\beta  \alpha } \Lambda ^n{}_{\beta },\nonumber\\
\left(\eta \sigma \Lambda ^k\right)^{\alpha }&\equiv&\eta _{l n} \left(\sigma ^{n k}\right)^{\beta  \alpha } \Lambda ^l{}_{\beta },\nonumber\\
\left(\eta \sigma \Lambda \Lambda ^k\right)^l&\equiv&\eta _{n o} \left(\sigma ^{o k}\right)^{\alpha  \beta } \Lambda ^n{}_{\alpha } \Lambda ^l{}_{\beta },\nonumber\\
\left(\sigma \Lambda \Lambda ^{k l}\right)^{n o}&\equiv&\left(\sigma ^{k l}\right)^{\alpha  \beta } \Lambda ^n{}_{\alpha } \Lambda ^o{}_{\beta }.
\end{eqnarray}

Following the strategy to construct a renormalizable action on the NAC superspace~\cite{Grisaru:2003fd}, we have to calculate the effective action of the deformed action ${\cal S}_{{\rm NC}}$.
In order to deal with the interacting vertices systematically, we choose the background field method~\cite{Gates:1983nr}.
Because there are many terms in the effective action, we adopt a short notation to represent the effective action. That is, we just list independent BFNC parameters and operators, where 
$\partial \partial$ denotes $\partial_k \partial_l$, and $D$ and $\bar{D}$ stand for $D_{\alpha}$ and $\bar{D}_{\dot{\beta}}$, respectively.
For instance, in ${\cal S}_{\rm NC}$ (eq.~(\ref{S_NC})) the terms at the second order of BFNC parameters (order of $\Lambda^2$), i.e. ${\cal S}_{\Lambda}(\Lambda^2)$ can be summarized as
\begin{itemize}
\item  5 BFNC parameters
\begin{equation}
\label{}
\Lambda ^{k l},\qquad \left(\sigma \Lambda \Lambda ^{k l}\right)^{n o},\qquad \eta ^{k l} \Lambda ^{n o},\qquad \epsilon ^{\alpha  \beta } \Lambda ^{k l},\qquad \epsilon ^{\alpha  \beta } \epsilon ^{\zeta  \iota } \Lambda ^k{}_{\beta } \Lambda ^l{}_{\iota }.
\end{equation}

\item   3 operators
\begin{equation}
\label{}
\partial \partial \Phi  \left(D^2 \Phi \right) \left(D^2 \Phi \right),\qquad
\partial \partial (D \Phi ) (D \Phi ) \left(D^2 \Phi \right),\qquad
\partial \partial \partial \partial \Phi ^+ \Phi ^+ \Phi ^+.
\end{equation}
\end{itemize}
Therefore,  the terms in the effective action of ${\cal S}_{\rm NC}$ (denoted by $\Gamma _{\rm 1st}$) at the order of $\Lambda^2$ are concisely represented as
\begin{itemize}
\item 14 BFNC parameters
\begin{eqnarray}
\label{gama1bfnc}
&&\left(\eta \sigma \Lambda \Lambda ^k\right)^l,\qquad \Lambda ^2,\qquad \Lambda ^{k l},\qquad \epsilon ^{\alpha  \beta } \left(\eta \sigma \Lambda \Lambda ^k\right)^l,\qquad \epsilon ^{\alpha  \beta } \Lambda ^{k l},\nonumber\\
&&\Lambda ^2 \eta ^{k l},\qquad \Lambda ^2 \epsilon ^{\alpha  \beta },\qquad \Lambda ^2 \epsilon ^{\dot{\alpha } \dot{\beta }},\qquad \Lambda ^2 \left(\bar{\sigma }^k\right)^{\dot{\alpha } \beta },\qquad \Lambda ^2 \eta ^{k l} \epsilon ^{\alpha  \beta },\nonumber\\
&&\Lambda ^{k l} \eta _{l n} \left(\bar{\sigma }^n\right)^{\dot{\alpha } \beta },\qquad \eta _{k l} \left(\bar{\sigma }^l\right)^{\dot{\alpha } \beta } \left(\eta \sigma \Lambda \Lambda ^k\right)^n,\qquad \epsilon ^{\alpha  \beta } \epsilon ^{\zeta  \iota } \Lambda ^k{}_{\beta } \Lambda ^l{}_{\iota },\nonumber\\
&&\epsilon ^{\alpha  \beta } \eta _{k l} \left(\sigma \Lambda ^{l n}\right)^{o \zeta } \Lambda ^k{}_{\beta }.
\end{eqnarray}

\item operators of point functions
\begin{itemize}

\item 4 operators of 2-point functions
\begin{equation}
\label{gama1op2}
\partial \partial \Phi  \left(D^2 \Phi \right),\qquad
\partial \partial (D \Phi ) (D \Phi ),\qquad
\partial \partial \left(D^2 \Phi \right) \Phi ^+,\qquad
\partial \partial \partial \partial \left(D^2 \Phi \right) \Phi ^+;
\end{equation}

\item 5 operators of 3-point functions
\begin{eqnarray}
\label{gama1op3}
&&\left(D^2 \Phi \right) \left(D^2 \Phi \right) \left(\bar{D}^2 \Phi ^+\right),\qquad
\partial (D \Phi ) \left(D^2 \Phi \right) \left(\bar{D} \Phi ^+\right),\qquad
\partial \partial \Phi  \left(D^2 \Phi \right) \Phi ^+,\nonumber\\
&&\partial \partial (D \Phi ) (D \Phi ) \Phi ^+,\qquad
\partial \partial \left(D^2 \Phi \right) \Phi ^+ \Phi ^+;
\end{eqnarray}

\item 5 operators of 4-point functions
\begin{eqnarray}
\label{gama1op4}
&&\left(D^2 \Phi \right) \left(D^2 \Phi \right) \left(\bar{D} \Phi ^+\right) \left(\bar{D} \Phi ^+\right),\qquad
\left(D^2 \Phi \right) \left(D^2 \Phi \right) \Phi ^+ \left(\bar{D}^2 \Phi ^+\right),\nonumber\\
&&\partial (D \Phi ) \left(D^2 \Phi \right) \Phi ^+ \left(\bar{D} \Phi ^+\right),\qquad
\partial \partial \Phi  \left(D^2 \Phi \right) \Phi ^+ \Phi ^+,\nonumber\\
&&\partial \partial (D \Phi ) (D \Phi ) \Phi ^+ \Phi ^+.
\end{eqnarray}

\end{itemize}
\end{itemize}

Because new BFNC parameters and operators appear in $\Gamma _{\rm 1st}$ (see eqs.~(\ref{gama1bfnc})-(\ref{gama1op4})), the terms composed of the new BFNC parameters and operators appeared  in $\Gamma _{\rm 1st}$ cannot be absorbed by ${\cal S}_{\rm NC}$. Thus, in accordance with the background field method~\cite{Gates:1983nr} we define a new action,
\begin{equation}
\label{action_S_1}
{\cal S}_{(1)}\equiv{\cal S}_{\rm WZ}+{\cal S}_{\Lambda }\left(\Lambda ^2\right)+\Gamma _{\rm 1st}\left(\Lambda ^2\right),
\end{equation}
and further derive the effective action of ${\cal S}_{(1)}$, which is denoted as $\Gamma _{\rm 2nd}\left(\Lambda ^2\right)$.
By comparing $\Gamma _{\rm 2nd}\left(\Lambda ^2\right)$ with ${\cal S}_{(1)}$, we find that the following BFNC parameters and operators are not contained in ${\cal S}_{(1)}$.
\begin{itemize}
\item 13 BFNC parameters
\begin{eqnarray}
\label{}
&&\sigma \Lambda \Lambda ,\qquad \Lambda ^2 \left(\sigma ^{k l}\right)^{\alpha  \beta },\qquad \sigma \Lambda \Lambda  \eta ^{k l},\qquad \sigma \Lambda \Lambda  \epsilon ^{\alpha  \beta },\qquad \sigma \Lambda \Lambda  \epsilon ^{\dot{\alpha } \dot{\beta }},\qquad\sigma \Lambda \Lambda  \left(\bar{\sigma }^k\right)^{\dot{\alpha } \beta },\nonumber\\
&& \epsilon ^{\alpha  \beta } \left(\eta \sigma \Lambda ^k\right)^{\zeta } \Lambda ^l{}_{\beta },\qquad \Lambda ^2 \eta ^{k l} \eta ^{n o},\qquad \Lambda ^{k l} \eta _{l n} \left(\sigma ^{n o}\right)^{\alpha  \beta },\qquad \sigma \Lambda \Lambda  \eta ^{k l} \epsilon ^{\alpha  \beta },\nonumber\\
&&\eta _{k l} \left(\bar{\sigma }^l\right)^{\dot{\alpha } \beta } \left(\eta \sigma \Lambda \Lambda ^n\right)^k,\qquad \epsilon ^{\alpha  \beta } \eta _{k l} \left(\eta \sigma \Lambda ^l\right)^{\zeta } \Lambda ^k{}_{\beta },\qquad \epsilon ^{\alpha  \beta } \epsilon ^{\zeta  \iota } \epsilon ^{k l n o} \eta _{n p} \eta _{o q} \Lambda ^p{}_{\beta } \Lambda ^q{}_{\iota }.
\end{eqnarray}

\item operators of point functions
\begin{itemize}

\item 1 operator of 1-point functions
\begin{equation}
\label{}
D^2 \Phi;
\end{equation}

\item 10 operators of 2-point functions
\begin{eqnarray}
\label{}
&&\Phi  \left(D^2 \Phi \right),\qquad
 (D \Phi ) (D \Phi ),\qquad
 \left(D^2 \Phi \right) \left(D^2 \Phi \right),\qquad
 \left(D^2 \Phi \right) \left(\bar{D}^2 \Phi ^+\right),\nonumber\\
&&\left(D^2 \Phi \right) \Phi ^+,\qquad
 \partial (D \Phi ) \left(\bar{D} \Phi ^+\right),\qquad
 \partial \partial \Phi  \Phi ^+,\qquad
 \partial \partial \left(D^2 \Phi \right) \left(D^2 \Phi \right),\nonumber\\
&&\partial \partial \Phi ^+ \Phi ^+,\qquad
 \partial \partial \partial \partial \Phi ^+ \Phi ^+;
\end{eqnarray}

\item 13 operators of 3-point functions
\begin{eqnarray}
\label{}
&&\Phi  \Phi  \left(D^2 \Phi \right),\qquad
 \Phi  (D \Phi ) (D \Phi ),\qquad
 \Phi  \left(D^2 \Phi \right) \left(D^2 \Phi \right),\qquad
 \Phi  \left(D^2 \Phi \right) \Phi ^+,\nonumber\\
&&(D \Phi ) (D \Phi ) \left(D^2 \Phi \right),\qquad
 (D \Phi ) (D \Phi ) \Phi ^+,\qquad
 \left(D^2 \Phi \right) \left(D^2 \Phi \right) \Phi ^+,
\nonumber\\
&& \left(D^2 \Phi \right) \left(\bar{D} \Phi ^+\right) \left(\bar{D} \Phi ^+\right),\qquad
\left(D^2 \Phi \right) \Phi ^+ \left(\bar{D}^2 \Phi ^+\right),\qquad
 \left(D^2 \Phi \right) \Phi ^+ \Phi ^+, \nonumber\\
&& \partial (D \Phi ) \Phi ^+ \left(\bar{D} \Phi ^+\right),\qquad
\partial \partial \Phi  \Phi ^+ \Phi ^+,\qquad
\partial \partial \Phi ^+ \Phi ^+ \Phi ^+;
\end{eqnarray}

\item 14 operators of 4-point functions
\begin{eqnarray}
\label{}
&&\Phi  \Phi  \left(D^2 \Phi \right) \left(D^2 \Phi \right),\qquad
 \Phi  \Phi  \left(D^2 \Phi \right) \Phi ^+,\qquad
 \Phi  (D \Phi ) (D \Phi ) \left(D^2 \Phi \right),
\nonumber\\
&& \Phi  (D \Phi ) (D \Phi ) \Phi ^+,\qquad
\Phi  \left(D^2 \Phi \right) \left(D^2 \Phi \right) \Phi ^+,\qquad
 \Phi  \left(D^2 \Phi \right) \Phi ^+ \Phi ^+,
 \nonumber\\
&& (D \Phi ) (D \Phi ) \left(D^2 \Phi \right) \Phi ^+,\qquad
(D \Phi ) (D \Phi ) \Phi ^+ \Phi ^+,\qquad
\left(D^2 \Phi \right) \Phi ^+ \left(\bar{D} \Phi ^+\right) \left(\bar{D} \Phi ^+\right),
\nonumber\\
&&  \left(D^2 \Phi \right) \Phi ^+ \Phi ^+ \left(\bar{D}^2 \Phi ^+\right),\qquad
 \left(D^2 \Phi \right) \Phi ^+ \Phi ^+ \Phi ^+,\qquad
 \partial (D \Phi ) \Phi ^+ \Phi ^+ \left(\bar{D} \Phi ^+\right),
\nonumber\\ 
&&\partial \partial \Phi  \Phi ^+ \Phi ^+ \Phi ^+,\qquad
 \partial \partial \Phi ^+ \Phi ^+ \Phi ^+ \Phi ^+;
\end{eqnarray}

\item 10 operators of 5-point functions
\begin{eqnarray}
\label{}
&&\Phi  \Phi  \left(D^2 \Phi \right) \left(D^2 \Phi \right) \Phi ^+,\qquad
 \Phi  \Phi  \left(D^2 \Phi \right) \Phi ^+ \Phi ^+,\qquad
 \Phi  (D \Phi ) (D \Phi ) \left(D^2 \Phi \right) \Phi ^+,\nonumber\\
&& \Phi  (D \Phi ) (D \Phi ) \Phi ^+ \Phi ^+,\qquad
\Phi  \left(D^2 \Phi \right) \Phi ^+ \Phi ^+ \Phi ^+,\qquad
 (D \Phi ) (D \Phi ) \Phi ^+ \Phi ^+ \Phi ^+,\nonumber\\
&& \left(D^2 \Phi \right) \Phi ^+ \Phi ^+ \left(\bar{D} \Phi ^+\right) \left(\bar{D} \Phi ^+\right),\qquad
 \left(D^2 \Phi \right) \Phi ^+ \Phi ^+ \Phi ^+ \left(\bar{D}^2 \Phi ^+\right),\nonumber\\
&&\partial (D \Phi ) \Phi ^+ \Phi ^+ \Phi ^+ \left(\bar{D} \Phi ^+\right),\qquad
 \partial \partial \Phi  \Phi ^+ \Phi ^+ \Phi ^+ \Phi ^+;
\end{eqnarray}

\item 2 operators of 6-point functions
\begin{eqnarray}
\label{}
\Phi  \Phi  \left(D^2 \Phi \right) \Phi ^+ \Phi ^+ \Phi ^+,\qquad
 \Phi  (D \Phi ) (D \Phi ) \Phi ^+ \Phi ^+ \Phi ^+.
\end{eqnarray}

\end{itemize}
\end{itemize}
The above result indicates that $\Gamma _{\rm 2nd}\left(\Lambda ^2\right)$ cannot be absorbed by ${\cal S}_{(1)}$.
So we have to continue the iteration and further define the following action,
\begin{equation}
\label{}
{\cal S}_{(2)}\equiv {\cal S}_{(1)}+\Gamma _{\rm 2nd}\left(\Lambda ^2\right),
\end{equation}
and use $\Gamma _{\rm 3rd}\left(\Lambda ^2\right)$ to represent its effective action at the order of $\Lambda ^2$.
By comparing $\Gamma _{\rm 3rd}\left(\Lambda ^2\right)$ with ${\cal S}_{(2)}$, we observe that only the following two independent BFNC parameters are not contained in ${\cal S}_{(2)}$,
\begin{equation} 
\label{}
\eta ^{k l} \left(\eta \sigma \Lambda \Lambda ^n\right)^o,\qquad \sigma \Lambda \Lambda  (\sigma ^{k l})^{\alpha  \beta },
\end{equation}
while all the operators in $\Gamma _{\rm 3rd}\left(\Lambda ^2\right)$ are covered by ${\cal S}_{(2)}$.

By now it is not necessary to continue the iteration process, but to analyze ${\cal S}_{\Lambda }\left(\Lambda ^2\right)$ and the three effective actions systematically. One important result obtained in our previous work~\cite{Miao:2013a} is the introduction of a $1/2$ supersymmetry invariant subset and its corresponding basis.
It is known that anyone of the four actions ${\cal S}_{\Lambda }\left(\Lambda ^2\right)$, $\Gamma _{\rm 1st}\left(\Lambda ^2\right)$, $\Gamma _{\rm 2nd}\left(\Lambda ^2\right)$, and $\Gamma _{\rm 3rd}\left(\Lambda ^2\right)$ can be separated into several classes, each of which as a close set contains the minimal number of terms and is invariant under the $1/2$ supersymmetry transformation. Any two classes have no common terms.
So we introduce the concept of $1/2$ supersymmetry invariant subsets and name every class as a $1/2$ supersymmetry invariant subset. The numbers of invariant subsets are 4, 17, 64, and 73 for the four actions, respectively. By comparing the invariant subsets of the four actions, we see that some subsets have same BFNC parameters and operators but different coefficients. After summing up all of the subsets of the four actions, we finally determine that there are 74 independent subsets in total, each of which is invariant under the $1/2$ supersymmetry transformation. In order to give the ingredients of an invariant subset provided by the four actions, we construct a new action,
\begin{equation}
\label{}
\Gamma\left(\Lambda ^2\right)={a_0} \,{\cal S}_{\Lambda }\left(\Lambda ^2\right)
+{a_1} \,\Gamma _{\rm 1st}\left(\Lambda ^2\right)
+{a_2}\, \Gamma _{\rm 2nd}\left(\Lambda ^2\right)
+{a_3}\, \Gamma _{\rm 3rd}\left(\Lambda ^2\right),
\end{equation}
where ${a_0}$, ${a_1}$, ${a_2}$, and ${a_3}$ are parameters which show intersections of invariant subsets of different actions.
All invariant subsets are given in ref.~\cite{Miao:2013a} and denoted by $f_i$, where $i=1,\cdots,74$. Thus, $\Gamma\left(\Lambda ^2\right)$ can be rewritten as
\begin{equation}
\label{}
\Gamma\left(\Lambda ^2\right)=\sum_{i=1}^{74} f_i.
\end{equation}

Further, based on the 74 invariant subsets $f_i$'s, we need to construct more general $1/2$ supersymmetry invariant subsets in order to deduce the one-loop renormalizable Wess-Zumino action on the BFNC superspace. This means that we only care about the combinations of BFNC parameters and operators rather than the coefficients of the terms in every invariant subset. The reason is that the coefficients merely represent the ratio of contribution to a combination from ${\cal S}_{\Lambda }\left(\Lambda ^2\right)$, $\Gamma _{\rm 1st}\left(\Lambda ^2\right)$, $\Gamma _{\rm 2nd}\left(\Lambda ^2\right)$, and $\Gamma _{\rm 3rd}\left(\Lambda ^2\right)$. By introducing parameters $x_{i,j}$, $y_{i,j}$, and $z_{i,j}$ to replace the original coefficients in every subset and again imposing the $1/2$ supersymmetry upon the generalized subset, where $i, j=1, 2, \cdots, 74$, we can have a generalized invariant subset containing those parameters that satisfy the constraints from the $1/2$ supersymmetry, $B_i$, which is given in ref.~\cite{Miao:2013a} and called a basis of an invariant subset.
Then, using the bases $B_i$'s rather than  the  four actions ${\cal S}_{\Lambda }\left(\Lambda ^2\right)$, $\Gamma _{\rm 1st}\left(\Lambda ^2\right)$, $\Gamma _{\rm 2nd}\left(\Lambda ^2\right)$, and $\Gamma _{\rm 3rd}\left(\Lambda ^2\right)$,  we define a general action ${\cal S}_{(3)}$ that is obviously invariant under the $1/2$ supersymmetry transformation,
\begin{equation}
\label{S_3}
{\cal S}_{(3)}={\cal S}_{\rm WZ}+\int d^8z\left(\sum _{i=1}^{74} B_i\right).
\end{equation}
By evaluating its effective action, $\Gamma _{\rm 4th}\left(\Lambda ^2\right)$, we observe that $\Gamma _{\rm 4th}\left(\Lambda ^2\right)$ no longer has new terms that do not exist in ${\cal S}_{(3)}$, i.e.  $\Gamma _{\rm 4th}\left(\Lambda ^2\right)$ can be absorbed completely by ${\cal S}_{(3)}$. Consequently,  ${\cal S}_{(3)}$ is the one-loop renormalizable BFNC Wess-Zumino action at the order of $\Lambda ^2$.

\section{Description of Idea and Treatment}

Now we turn to the construction of an all-loop renormalizable Wess-Zumino action on the BFNC superspace. In this section we briefly describe our main idea of investigation as follows.
A natural extension of the one-loop renormalizable BFNC Wess-Zumino action is an all-loop renormalizable one. As we know, ${\cal S}_{(3)}$ (eq.~(\ref{S_3})) with a real mass and a real interacting constant has only one $1/2$ supersymmetry but no global $U(1)$ symmetries. According to the realization of one- and all-loop renormalizable NAC Wess-Zumino actions~\cite{Britto:2003aj, Britto:2003kg, Romagnoni:2003xt},  
we have to modify ${\cal S}_{(3)}$ by generalizing the mass and interacting constant to complex numbers and then introduce two global $U(1)$ symmetries, i.e. the $U(1)_{\rm R}$ $R$-symmetry and $U(1)_{\Phi}$ flavor symmetry in the modified formulation of ${\cal S}_{(3)}$.
That is, by requiring the one-loop renormalizable Wess-Zumino action ${\cal S}_{(3)}$~\cite{Miao:2013a} to possess the $U(1)_{\rm R}$ $R$-symmetry and $U(1)_{\Phi}$ flavor symmetry, we give\footnote{Note that this is just an intermediate step. ${\cal S}^{\prime}_{(3)}$ is not a new BFNC Wess-Zumino action that is renormalizable at all loops. For the new BFNC Wess-Zumino action, see eq.~(\ref{finalaction}) or eq.~(\ref{alternativeaction}).}  a modified action ${\cal S}^{\prime}_{(3)}$ (see eq.~(\ref{action_1})) that has not only the $1/2$ supersymmetry but also the two $U(1)$ symmetries.
In the usual perturbation expansion to any order, one then tries to determine the effective action of ${\cal S}^{\prime}_{(3)}$ in terms of Feynman graphs. However,
we find that when the order of perturbation expansion increases, it is too complicated to be performed to calculate such an effective action for ${\cal S}^{\prime}_{(3)}$ that contains a so large number of terms. Consequently, we have to open a new road for ourselves. The key factor relies on symmetries. 
As ${\cal S}^{\prime}_{(3)}$ has two $U(1)$ symmetries and one $1/2$ supersymmetry, according to the background field method~\cite{Gates:1983nr}, the effective action of ${\cal S}^{\prime}_{(3)}$ would certainly have such symmetries even if it has not yet been derived.
This implies that we shall obtain the effective action of ${\cal S}^{\prime}_{(3)}$ by just using these symmetries.

Based on the above idea, we then demonstrate briefly our specific treatment of investigation. We begin with constructing all possible divergent operators in the effective action of ${\cal S}^{\prime}_{(3)}$ by using the two $U(1)$ symmetries.
At this stage, only the number of derivatives and the number of superfields can be determined, while the distributions of derivatives and superfields, i.e. the positions of derivatives and superfields in an allowed divergent operator cannot be fixed.
According to the general requirement of renormalization in the background field method~\cite{Gates:1983nr, Dimitrijevic:2010yv}, if the effective action of an action can be absorbed by the action itself, then such an action is regarded to be renormalizable; if not, an iteration should be processed. By adding the effective action to the action, one then gives a combined action. If the effective action of the combined action can be absorbed by the combined one, then such a combined action is regarded to be renormalizable; if not, such an iterative procedure should be continued until to the stage that an effective action is absorbed by its corresponding action. Therefore, 
as a preliminary analysis, we determine whether ${\cal S}^{\prime}_{(3)}$ contains a term that has the same number of derivatives and the same number of superfields as that of a possible divergent operator in the effective action of ${\cal S}^{\prime}_{(3)}$.  
Then, in a more detailed analysis, for every class of divergent operator forms, we check whether ${\cal S}^{\prime}_{(3)}$ contains such a term that has not only  the same numbers of derivatives and superfields but also the same distributions of derivatives and superfields as that of an allowed divergent operator.
According to the negative and positive situations, we divide all divergent operators into two parts.
The first part, denoted as the first, second and third classes (see eqs.~(\ref{part_1})-(\ref{part_3})), has no corresponding terms in ${\cal S}^{\prime}_{(3)}$, while the second part,  denoted as the fourth class (see eqs.~(\ref{part_4_1})-(\ref{part_4_6})), has corresponding terms in ${\cal S}^{\prime}_{(3)}$. Further adding\footnote{See our detailed analyses in sections 4-8.} specific terms from the third and fourth classes of divergent operators to ${\cal S}^{\prime}_{(3)}$ after considering all possible combinations of allowed BFNC parameters and divergent operators, we can finally decide a new BFNC Wess-Zumino action (eq.~(\ref{finalaction}) or eq.~(\ref{alternativeaction})) that is renormalizable at all loops. The remaining paragraphs of this section leave to a concise description of such a process.

Although the part of the effective action constructed from the first part of divergent operators cannot be absorbed by ${\cal S}^{\prime}_{(3)}$, 
not all of its divergent operators have to be considered.
By the requirement of $1/2$ supersymmetry and the fact that some divergent operators can be changed to total derivatives,
we can get rid of the first and second classes of divergent operators in the first part.
Thus only the third class (see eq.~(\ref{part_3})) remains.
In addition, after considering all allowed ways for acting covariant derivatives on superfields, we can give the $1/2$ supersymmetric forms (eqs.~(\ref{action_3rd_1}) and (\ref{action_3rd_2})) by using the third class.

For the divergent operators in the second part (fourth class), because ${\cal S}^{\prime}_{(3)}$ contains a lot of terms, one naturally wants to ask whether ${\cal S}^{\prime}_{(3)}$ covers this part of the effective action constructed by using these divergent operators.
To answer the question, we have to give this part of the effective action by using the fourth class of divergent operators.
At first we work out all allowed BFNC parameters, determine the combinations of BFNC parameters and divergent operators, and then  deduce the distributions of derivatives on the superfields by using the $1/2$ supersymmetry.
Although the derivation of an effective action from divergent operators is very complicated in general, we can give the effective action of ${\cal S}^{\prime}_{(3)}$ in light of our specific observation that ${\cal S}^{\prime}_{(3)}$ covers a large part of the effective action, i.e. 
only a small number of terms (see eq.~(\ref{action_4th})) are not contained in ${\cal S}^{\prime}_{(3)}$.

As a consequence, the terms in eqs.~(\ref{action_3rd_1}), (\ref{action_3rd_2}), and (\ref{action_4th}) are not covered by ${\cal S}^{\prime}_{(3)}$.
This means that if these terms appear in the effective action of ${\cal S}^{\prime}_{(3)}$ at a certain order of perturbation,
they cannot be absorbed by ${\cal S}^{\prime}_{(3)}$, which gives rise to the result that ${\cal S}^{\prime}_{(3)}$ is not renormalizable.
In order to determine whether the terms in eqs.~(\ref{action_3rd_1}), (\ref{action_3rd_2}), and (\ref{action_4th}) appear in the effective action of ${\cal S}^{\prime}_{(3)}$, the usual way is to calculate the effective action to all orders in perturbation theory. 
Without applying the usual direct perturbative investigation which is unsuitable to our case as analyzed already, using our specific treatment that is based on the three symmetries mentioned above, we impose these terms directly upon ${\cal S}^{\prime}_{(3)}$ and find that the combined  action is renormalizable at all loops.

\section{The Modified One-Loop Renormalizable Wess-Zumino Action}

We need to set up the symmetry charge for the BFNC parameter ${\Lambda ^k}{}_{\alpha }$, see Table~\ref{symmetry}. 
\begin{table}[t]
\centering
\begin{tabular}{|c|c|c|c|c|c|c|c|}
\hline
  &  ${\rm dim}$& $U(1)_{\rm R}$ & $U(1)_{\Phi}$ &  & ${\rm dim}$& $U(1)_{\rm R}$ & $U(1)_{\Phi}$\\
\hline
$m$ & 1 & 0 & -2 & $m^*$ & 1 & 0 & 2\\
\hline
$g$&0&-1&-3&$g^*$&0&1&3\\
\hline
$ (\Lambda ^k{}_{\alpha } )^2$&-3&2&0&V&-5&2&0\\
\hline
$d^4\theta$ &2&0&0&$\theta ^4$&-2&0&0\\
\hline
$\Phi$ &1&1&1&$\Phi ^+$&1&-1&-1\\
\hline
$D_{\alpha}$&$\frac{1}{2}$&-1&0&$\bar{D}_{\dot{\alpha }}$&$\frac{1}{2}$&1&0\\
\hline
${D}^2$&1&-2&0&$\bar{D}^2$&1&2&0\\
\hline
$\partial _k$&1&0&0&$d^4 x$&-4&0&0\\
\hline
\end{tabular}
\caption{\small
Mass dimensions and symmetry charges of parameters and operators.
\label{symmetry}
}
\end{table}
In order to modify ${\cal S}_{(3)}$ (see ref.~\cite{Miao:2013a}) to be such a  form that has the additional $U(1)_{\rm R}$ $R$-symmetry and $U(1)_{\Phi}$ flavor symmetry, we generalize the original real mass $m$ and real interacting constant $g$ to complex numbers, where $m^*$ and $g^*$ represent complex conjugate to $m$ and $g$, respectively, and define the parameters $x_{i,j}$, $y_{i,j}$, and $z_{i,j}$ in ${\cal S}_{(3)}$ to have no mass dimensions or $U(1)$ charges. The modified action can thus be given by following the two-step manipulation of ${\cal S}_{(3)}$.

{\em The first step}: For the undeformed part of ${\cal S}_{(3)}$, which corresponds to the ordinary Wess-Zumino action, we replace $m$ and $g$ related to antichiral fields by $m^*$ and $g^*$, respectively, and thus give the result as follows,
\begin{eqnarray}
\label{WZ_2_superfield_form}
{\cal S}^{\prime}_{\rm WZ}&=&\int d^8z\left\{\Phi ^+\Phi -\frac{m}{8}\Phi  \left(\frac{D^2}{\square }\Phi \right)-\frac{m^*}{8}\Phi ^+ \left(\frac{\bar{D}^2}{\square }\Phi ^+\right)\right.\nonumber\\
&&\hspace{15mm}\left.-\frac{g}{12}\Phi  \Phi \left(\frac{D^2}{\square }\Phi \right)-\frac{g^*}{12}\Phi ^+ \Phi ^+ \left(\frac{\bar{D}^2}{\square }\Phi ^+\right)\right\}.
\end{eqnarray}

{\em The second step}: For  the deformed part of ${\cal S}_{(3)}$,  to each base denoted by $\int d^8 z \,X$  we at first multiply $X$ by $m^{b_1} (m^*)^{b_2} g^{b_3} (g^*)^{b_4}$, where $b_1$, $b_2$, $b_3$, and $b_4$ are parameters.
Then considering the requirement of the $U(1)_{\rm R}$ $R$-symmetry and $U(1)_{\Phi}$ flavor symmetry, we observe that the mass dimension,  the $U(1)_{\rm R}$ charge, and the $U(1)_{\Phi}$ charge of $\int d^8 z\,m^{b_1} (m^*)^{b_2} g^{b_3} (g^*)^{b_4} \,X$ should be zero, which gives rise to constraints on the four parameters. 
By solving these constraints, we can fix the parameters $b_1$, $b_2$, $b_3$, and $b_4$ and thus obtain ${\cal S}^{\prime}_{(3)}$.

We take $B_{29}$ as an example to explain how to modify the coefficient of each base in the action ${\cal S}_{(3)}$ and then derive the corresponding base in ${\cal S}^{\prime}_{(3)}$.  $B_{29}$ has the form~\cite{Miao:2013a},
\begin{eqnarray}
\label{basis_example}
B_{29}&=&\left(-2 {x_{29,3}}+2 {x_{29,4}}\right) \epsilon ^{\alpha  \beta } \Lambda ^{k l} \theta ^4 \left(D_{\beta } \Phi \right) \partial _l\partial _k\left(D_{\alpha } \Phi \right) \left(D^2 \Phi \right)\nonumber\\
&&+\left(-6 {x_{29,3}}+4 {x_{29,4}}+2 {x_{29,5}}\right) \epsilon ^{\alpha  \beta } \epsilon ^{\zeta  \iota } \Lambda ^k{}_{\beta } \Lambda ^l{}_{\iota } \theta ^4 \partial _k\left(D_{\alpha } \Phi \right) \partial _l\left(D_{\zeta } \Phi \right) \left(D^2 \Phi \right)\nonumber\\
&&+{x_{29,3}} \Lambda ^{k l} \theta ^4 \Phi  \partial _k\left(D^2 \Phi \right) \partial _l\left(D^2 \Phi \right)\nonumber\\
&&+{x_{29,4}} \Lambda ^{k l} \theta ^4 \Phi  \left(D^2 \Phi \right) \partial _l\partial _k\left(D^2 \Phi \right)\nonumber\\
&&+{x_{29,5}} \epsilon ^{\alpha  \beta } \Lambda ^{k l} \theta ^4 \partial _k\left(D_{\alpha } \Phi \right) \partial _l\left(D_{\beta } \Phi \right) \left(D^2 \Phi \right).
\end{eqnarray}
Following the second step, where $X$ now represents $B_{29}$, we can compute $b_1=b_2=b_4=0$ and $b_3=1$. Thus, the modified $B_{29}$ takes the form $B_{29}^{\prime}=g B_{29}$. After dealing with the other bases in ${\cal S}_{(3)}$ in the same way, we deduce from ${\cal S}_{(3)}$ the modified action,
\begin{equation}
\label{action_1}
{\cal S}^{\prime}_{(3)}={\cal S}^{\prime}_{\rm WZ}+\int d^8z\left(\sum _{i=1}^{74} c_i B_i\right), 
\end{equation} 
where $B_i$'s ($i=1, \cdots, 74$) have been listed in Appendix B of ref.~\cite{Miao:2013a},  and $c_i$'s denote the additional coefficients we have calculated according to {\em The second step} mentioned above. The coefficients are listed in the Appendix, see eq.~(\ref{coefficients_c}). For instance, the coefficient related to $B_{29}$ is $c_{29}=g$.
As a result, ${\cal S}^{\prime}_{(3)}$ has the $U(1)_{\rm R}$ $R$-symmetry, $U(1)_{\Phi}$ flavor symmetry, and $1/2$ supersymmetry. This modified action will play a fundamental role in our construction of  an all-loop renormalizable Wess-Zumino model on the BFNC superspace.

\section{Constraints on Divergent Operators from Symmetries}

A general term in an effective action 
can be written in the following form~\cite{Romagnoni:2003xt},
\begin{equation}
\label{effective_action}
\Gamma =\int d^4x~ \lambda~  {\mathcal O},
\end{equation}
where $ {\mathcal O}$ denotes a general divergent operator that includes $d^4\theta$, and the parameter $\lambda$ is defined up to a coefficient as
\begin{equation}
\label{constant}
\lambda \sim \Lambda_{UV} ^dg^{x-R}\left(g^*\right)^x\left(\frac{m}{\Lambda_{UV} }\right)^y\left(\frac{{m^*}}{\Lambda_{UV} }\right)^{y+\frac{S-3R}{2}}.
\end{equation}
As to the quantities in the above equation, we give an explanation.
$\Lambda_{UV}$ is ultraviolet cutoff, and $x$ and $y$ are non-negative integers. 
Moreover,  $d$, $R$, and $S$ represent, for the parameter $\lambda$, its mass dimension, its $U(1)_{\rm R}$ $R$-symmetry charge, and its  $U(1)_{\Phi}$ flavor symmetry charge, respectively, which can be obtained from Table~\ref{symmetry}.

By considering the $D$ algebraic relations~\cite{Miao:2013a} we can determine the most general operator  ${\mathcal O}$, 
\begin{equation}
\label{operator}
 {\mathcal O}=d^4\theta  (D^2)^{\gamma }  (\bar{D}^2)^{\delta } (\partial D  \bar{D})^{\eta }(\partial\partial)^{\zeta }V^{\rho }\Phi ^{\alpha }  (\Phi ^+)^{\beta },
\end{equation}
where $\gamma$, $\delta$, $\eta$, $\zeta$, $\rho$, $\alpha$, and $\beta$ are non-negative  integers, 
$\partial D \bar{D}$ stands for $\partial _k D_{\alpha }\bar{D}_{\dot{\beta }}$, $\partial\partial$ for $\partial _k\partial _l$, and $V$  for $(\Lambda ^k{}_{\alpha } )^2\,\theta ^4$.
The dimension, $U(1)_{\rm R}$ charge, and $U(1)_{\Phi}$ charge of ${\mathcal O}$ can be obtained from Table~\ref{symmetry}, which are given respectively as follows,
\begin{equation}
\label{dim_O}
2+\alpha +\beta +\gamma +\delta +2 \zeta +2 \eta -5 \rho,
\end{equation}
\begin{equation}
\label{R_charge}
\alpha -\beta +2 (-\gamma +\delta +\rho ),
\end{equation}
\begin{equation}
\label{F_charge}
\alpha -\beta.
\end{equation}

The dimension of the general term $\Gamma$ 
in eq.~(\ref{effective_action}) is the sum of the dimensions of $d^4 x$, $\lambda$, and ${\mathcal O}$. 
As the dimension of $\Gamma$ is zero and the dimension of $d^4 x$ is $-4$ by definition, so the sum of the dimensions of $\lambda$ and ${\mathcal O}$ is $4$.

Because $\Gamma$ 
is invariant under the $U(1)_{\rm R}$ $R$-transformation and $U(1)_{\Phi}$ flavor transformation, the $U(1)_{\rm R}$ and $U(1)_{\Phi}$ charges of $\Gamma$ are zero. 
We note that the symmetry charge of $d^4 x$ is defined to be zero, so the symmetry charge of $\Gamma$  is the sum of symmetry charges of $\lambda$ and ${\mathcal O}$.

Using eqs.~(\ref{dim_O}), (\ref{R_charge}), and (\ref{F_charge}), we deduce for $\lambda$ its dimension $d$, $U(1)_{\rm R}$ charge $R$, and $U(1)_{\Phi}$ charge $S$ as follows,
\begin{eqnarray}
\label{dRS}
d&=&2-\alpha -\beta -\gamma -\delta -2 \zeta -2 \eta +5 \rho, \nonumber\\
R&=&-\alpha +\beta +2 \gamma -2 \delta -2 \rho, \nonumber\\
S&=&-\alpha +\beta.
\end{eqnarray}

From eq.~(\ref{constant}) we can read off the power of $\Lambda_{UV}$ denoted as $P$,
\begin{equation}
\label{eq_P}
P=d+\frac{3 R}{2}-\frac{S}{2}-2 y.
\end{equation}
After substituting eq.~(\ref{dRS}) into eq.~(\ref{eq_P}) we obtain
\begin{equation}
\label{condition P1}
P=2-2 y-2 \alpha +2 \gamma -4 \delta -2 \zeta -2 \eta +2 \rho.
\end{equation}

From the definition of $\lambda$, we get $\lambda \propto (\Lambda _{{UV}})^P$. 
As $\Lambda_{{UV}}$ is  ultraviolet cutoff, it will be infinity. 
We thus conclude that $\Gamma$ 
(see eq.~(\ref{effective_action})) is divergent only when
\begin{equation}
\label{condition P2}
P\geq 0.
\end{equation}

Because we only care about the divergent part in an effective action, we impose the constraint eq.~(\ref{condition P2}) upon the possible forms of the parameter $\lambda$ and operator ${\mathcal O}$.

Besides eq.~(\ref{condition P2}), we have to consider the constraints which are related to the $D$ algebraic relations,
\begin{equation}
\label{condition D}
\gamma \leq  \alpha -\eta, \qquad \delta \leq \beta -\eta.
\end{equation}
The reason is now stated. 
If there is one $\partial D  \bar{D}$ in ${\mathcal O}$, then we must act $D$ on $\Phi$ and act $\bar{D}$ on $\Phi ^+$ separately in an effective action.
If not, first assuming that $D$ and $\bar{D}$ act on the same superfield, that is, $(\partial D  \bar{D} \Phi ^+)\Phi$, and 
then using $D$ algebraic relations, we get $(\partial \partial \Phi^+)\Phi$.
This contradicts the initial assumption that there is one $\partial D  \bar{D}$ in ${\mathcal O}$.
Taking into account this point, we can act at most two $D$'s on one $\Phi$ and two $\bar {D}$'s on one $\Phi ^+$, and thus give  the constraints eq.~(\ref{condition D}).

In addition, the other constraints emerge from the requirement that the powers of the factors in ${\mathcal O}$ are non-negative integers,
\begin{eqnarray}
\label{condition non 0}
&&\gamma \geq 0,\qquad \delta \geq 0,\qquad \alpha \geq 0,\qquad \eta \geq 0,\qquad \rho \geq 0,\nonumber\\
&&\zeta \geq 0,\qquad \beta \geq 0.
\end{eqnarray}

Next recall that we have defined $V\equiv (\Lambda ^k{}_{\alpha } )^2\,\theta ^4$ and used $\rho$ to represent the number of $V$ in ${\mathcal O}$, cf. eq.~(\ref{operator}). If our approximation is up to the second order of BFNC parameters $\Lambda ^k{}_{\alpha }$'s, we  have another constraint on  $\rho$,
\begin{equation}
\label{condition nc}
\rho =1.
\end{equation}

At last, the parameters $m$, $m^*$, $g$, and $g^*$ enter an effective action through vertices and propagators, and the number of the parameters  should be non-negative integers, so we have the constraints on the powers of these quantities in $\lambda$. By expanding $\lambda$ and collecting $m$, $m^*$, $g$, and $g^*$ in eq.~(\ref{constant}), we can read off the constraints on their powers,
\begin{eqnarray}
\label{condition int}
&&x+\alpha -\beta -2 \gamma +2 \delta +2 \rho \geq 0,\nonumber\\
&&y+\alpha -\beta -3 \gamma +3 \delta +3 \rho \geq 0,\nonumber\\
&&y\geq 0,\qquad x\geq 0.
\end{eqnarray}

By now we obtain all of the constraints eqs.~(\ref{condition P1})-(\ref{condition int}). 
They are just linear equations and can be solved easily.

\section{Divergent Operators}

By solving the constraints eqs.~(\ref{condition P1})-(\ref{condition int}), we obtain many solutions which correspond to possible forms of divergent operators ${\mathcal O}$. Through determining  whether such forms can be covered by ${\cal S}^{\prime}_{(3)}$ or not, in other words, if ${\cal S}^{\prime}_{(3)}$ contains the terms that have the same numbers of derivatives and superfields as that of the possible forms of ${\mathcal O}$,  we classify the solutions into two parts.

The first part corresponds to the case that ${\cal S}^{\prime}_{(3)}$ does not contain the following three classes of the forms of ${\mathcal O}$.

\begin{itemize}

\item 
The first class of divergent operators has twelve different forms,
\begin{eqnarray}
\label{part_1}
&&\Phi ,\qquad 
\Phi  \Phi ,\qquad 
\Phi  \Phi ^+,\qquad 
\Phi  \Phi ^+ \Phi ^+,\qquad
\Phi  \Phi ^+ \Phi ^+ \Phi ^+,\qquad\nonumber\\
&&\Phi  \Phi ^+ \Phi ^+ \Phi ^+ \Phi ^+,\qquad
\Phi  \Phi ^+ \Phi ^+ \Phi ^+ \Phi ^+ \Phi ^+,\qquad
\Phi  \Phi  \Phi ^+,\nonumber\\
&&\Phi  \Phi  \Phi ^+ \Phi ^+,\qquad 
\Phi  \Phi  \Phi ^+ \Phi ^+ \Phi ^+,\qquad
\Phi  \Phi  \Phi ^+ \Phi ^+ \Phi ^+ \Phi ^+,\nonumber\\
&&\Phi  \Phi  \Phi ^+ \Phi ^+ \Phi ^+ \Phi ^+ \Phi ^+;
\end{eqnarray}

\item  
The second class of divergent operators has five different forms,
\begin{eqnarray}
\label{part_2}
&&\partial\partial \Phi ,\qquad \partial\partial \Phi ^+,\qquad \partial\partial\partial\partial \Phi ^+,\qquad D^2 \partial\partial \Phi ,\qquad D^2 \partial\partial\partial\partial\Phi;
\end{eqnarray}

\item 
The third class of divergent operators has eleven different forms,
\begin{eqnarray}
\label{part_3}
&&\Phi ^+,\qquad 
\Phi ^+ \Phi ^+,\qquad 
\Phi ^+ \Phi ^+ \Phi ^+,\qquad 
\Phi ^+ \Phi ^+ \Phi ^+ \Phi ^+,\nonumber\\
&&\Phi ^+ \Phi ^+ \Phi ^+ \Phi ^+ \Phi ^+,\qquad
\bar{D}^2 \Phi ^+,\qquad 
\bar{D}^2 \Phi ^+ \Phi ^+,\nonumber\\
&&\bar{D}^2 \Phi ^+ \Phi ^+ \Phi ^+,\qquad
\bar{D}^2 \Phi ^+ \Phi ^+ \Phi ^+ \Phi ^+,\nonumber\\
&&\bar{D}^2 \Phi ^+ \Phi ^+ \Phi ^+ \Phi ^+ \Phi ^+,\qquad 
\bar{D}^2 \Phi ^+ \Phi ^+ \Phi ^+ \Phi ^+ \Phi ^+ \Phi ^+.
\end{eqnarray}

\end{itemize}

We note that the operators in the first class are excluded by the $1/2$ supersymmetry invariance and those in the second class are excluded by the property that  
they are related merely to total derivatives after the spacetime integration.
So, only the operators in the third class can be used to construct  $1/2$ supersymmetry invariants.

The second part corresponds to the case that  ${\cal S}^{\prime}_{(3)}$ contains the following 1- to 6-point functions of ${\mathcal O}$ that are denoted as the fourth class of the forms of ${\mathcal O}$.

\begin{itemize}

\item The 1-point function has only one form,
\begin{equation}
\label{part_4_1}
D^2 \Phi;
\end{equation}

\item The 2-point function has twelve different forms,
\begin{eqnarray}
\label{part_4_2}
&&D^2 \Phi  \Phi ,\qquad 
D^2 D^2 \Phi  \Phi ,\qquad 
\partial\partial \Phi ^+ \Phi ^+,\qquad 
\partial\partial\partial\partial \Phi ^+ \Phi ^+,\nonumber\\
&&D^2 \partial\partial\Phi  \Phi ,\qquad 
D^2 \Phi  \Phi ^+,\qquad 
D^2 D^2 \partial\partial \Phi  \Phi ,\qquad 
\partial\partial \Phi  \Phi ^+, \nonumber\\
&&\partial D  \bar{D} \Phi  \Phi ^+,\qquad 
D^2 \partial\partial \Phi  \Phi ^+,\qquad 
D^2 \partial\partial\partial\partial \Phi  \Phi ^+, \qquad D^2 \overline{D}^2 \Phi  \Phi ^+;
\end{eqnarray}

\item The 3-point function has fifteen different forms,
\begin{eqnarray}
\label{part_4_3}
&&D^2 \Phi  \Phi  \Phi ,\qquad 
D^2 D^2 \Phi  \Phi  \Phi ,\qquad 
\partial\partial\Phi ^+ \Phi ^+ \Phi ^+,\nonumber\\
&&\partial\partial\partial\partial \Phi ^+ \Phi ^+ \Phi ^+,\qquad 
D^2 \Phi  \Phi ^+ \Phi ^+,\qquad 
D^2 \Phi  \Phi  \Phi ^+,\nonumber\\
&&D^2 D^2 \partial\partial \Phi  \Phi  \Phi ,\qquad
D^2 D^2 \Phi  \Phi  \Phi ^+,\qquad 
\partial\partial\Phi  \Phi ^+ \Phi ^+,\nonumber\\
&&\partial D  \bar{D}\Phi  \Phi ^+ \Phi ^+,\qquad 
D^2 \partial\partial \Phi  \Phi ^+ \Phi ^+,\qquad 
D^2 \partial\partial \Phi  \Phi  \Phi ^+,\nonumber\\
&&D^2 \partial D  \bar{D}\Phi  \Phi  \Phi ^+,\qquad 
D^2 \overline{D}^2 \Phi  \Phi ^+ \Phi ^+,\qquad
D^2 D^2 \overline{D}^2 \Phi  \Phi  \Phi ^+;
\end{eqnarray}

\item The 4-point function has twelve different forms,
\begin{eqnarray}
\label{part_4_4}
&&D^2 D^2 \Phi  \Phi  \Phi  \Phi ,\qquad 
\partial\partial\Phi ^+ \Phi ^+ \Phi ^+ \Phi ^+,\qquad 
D^2 \Phi  \Phi ^+ \Phi ^+ \Phi ^+,\nonumber\\
&&D^2 \Phi  \Phi  \Phi ^+ \Phi ^+,\qquad 
D^2 \Phi  \Phi  \Phi  \Phi ^+,\qquad 
D^2 D^2 \Phi  \Phi  \Phi  \Phi ^+,\nonumber\\
&&\partial\partial \Phi  \Phi ^+ \Phi ^+ \Phi ^+,\qquad
\partial D  \bar{D} \Phi  \Phi ^+ \Phi ^+ \Phi ^+, \qquad D^2 \partial\partial \Phi  \Phi  \Phi ^+ \Phi ^+, \nonumber\\
&& D^2 \partial D  \bar{D} \Phi  \Phi  \Phi ^+ \Phi ^+,\qquad
D^2 \overline{D}^2 \Phi  \Phi ^+ \Phi ^+ \Phi ^+,\qquad
D^2 D^2 \overline{D}^2 \Phi  \Phi  \Phi ^+ \Phi ^+;
\end{eqnarray}

\item The 5-point function has six different forms,
\begin{eqnarray}
\label{part_4_5}
&&D^2 \Phi  \Phi  \Phi ^+ \Phi ^+ \Phi ^+,\qquad 
D^2 \Phi  \Phi  \Phi  \Phi ^+ \Phi ^+,\qquad D^2 D^2 \Phi  \Phi  \Phi  \Phi  \Phi ^+, \nonumber\\
&& 
\partial\partial\Phi  \Phi ^+ \Phi ^+ \Phi ^+ \Phi ^+, \qquad  \partial D  \bar{D}\Phi  \Phi ^+ \Phi ^+ \Phi ^+ \Phi ^+,\qquad 
D^2 \overline{D}^2 \Phi  \Phi ^+ \Phi ^+ \Phi ^+ \Phi ^+;
\end{eqnarray}

\item The 6-point function has only one form,
\begin{eqnarray}
\label{part_4_6}
D^2 \Phi  \Phi  \Phi  \Phi ^+ \Phi ^+ \Phi ^+.
\end{eqnarray}

\end{itemize}

\section{BFNC Parameters}

Besides the various divergent operators given in the above section, we now determine allowed BFNC parameters because 
a general term in an effective action is a combination of allowed divergent operators and BFNC parameters, and such a combination has no free Bosonic and Fermionic indices and satisfies the $U(1)_{\rm R}$ $R$-symmetry, $U(1)_{\Phi}$ flavor symmetry, and $1/2$ supersymmetry. Here we list various possible forms of BFNC parameters
by using the following symbols,
\begin{eqnarray}
\label{all_symbol}
& & \epsilon _{\alpha  \beta }, \qquad \epsilon ^{\alpha  \beta },\qquad \epsilon _{\dot{\alpha }\dot{\beta }},\qquad \epsilon ^{\dot{\alpha}\dot{\beta }},\qquad \eta _{k l}, \qquad \eta ^{k l},\nonumber \\ 
& &  \epsilon ^{{klmn}}, \qquad  (\sigma ^{k l} )^{\alpha  \beta },\qquad  (\bar{\sigma }^k )^{\dot{\alpha }\beta },\qquad  \Lambda^k{}_{\alpha},
\end{eqnarray}
and 
classify the related BFNC parameters by their Bosonic indices $k$, $l$, $\cdots$ and Fermionic ones $\alpha$, $\beta$,  $\cdots$.

\begin{itemize}

\item The class without any indices contains two different forms,
\begin{equation}
\label{con_b0}
\Lambda ^2,\qquad \sigma \Lambda \Lambda;
\end{equation}

\item The class with two Bosonic indices $k$ and  $l$ contains four different forms,
\begin{equation}
\label{con_b2}
(\eta \sigma \Lambda \Lambda ^k)^l,\qquad \Lambda ^{k l},\qquad \Lambda ^2 \eta ^{k l},\qquad \sigma \Lambda \Lambda  \eta ^{k l};
\end{equation}

\item The class with two Fermionic indices $\alpha$ and $\beta$ contains two different forms,
\begin{equation}
\label{con_f2}
\Lambda ^2 \epsilon ^{\alpha  \beta },\qquad 
\sigma \Lambda \Lambda  \epsilon ^{\alpha  \beta };
\end{equation}

\item The class with two Fermionic indices $\dot{\alpha}$ and $\dot{\beta}$ contains two different forms,
\begin{equation}
\label{con_f2_dot}
\Lambda ^2 \epsilon ^{\dot{\alpha } \dot{\beta }},\qquad 
\sigma \Lambda \Lambda  \epsilon ^{\dot{\alpha } \dot{\beta }};
\end{equation}

\item The class with one Bosonic index $k$ and two Fermionic indices $\dot{\alpha }$ and $\beta$ contains five different forms,
\begin{eqnarray}
\label{con_b1_f2}
& &\Lambda ^2 (\bar{\sigma }^k)^{\dot{\alpha } \beta },\qquad 
\sigma \Lambda \Lambda  (\bar{\sigma }^k)^{\dot{\alpha } \beta },\qquad 
\Lambda ^{k l} \eta _{l n} (\bar{\sigma }^n)^{\dot{\alpha } \beta }, \nonumber\\
& &\eta _{n l} (\bar{\sigma }^l)^{\dot{\alpha } \beta } (\eta \sigma \Lambda \Lambda ^n)^k,\qquad
\eta _{n l} (\bar{\sigma }^l)^{\dot{\alpha } \beta } (\eta \sigma \Lambda \Lambda ^k)^n;
\end{eqnarray}

\item The class with four Bosonic indices $k$, $l$, $n$, and $o$ contains five different forms,
\begin{eqnarray}
\label{con_b4}
\Lambda ^2 \eta ^{k l} \eta ^{n o},\qquad 
\sigma \Lambda \Lambda  \eta ^{k l} \eta ^{n o},\qquad 
\eta ^{k l} \Lambda ^{n o},\qquad
(\sigma \Lambda \Lambda ^{k l})^{n o},\qquad 
\eta ^{k l} (\eta \sigma \Lambda \Lambda ^n)^o;
\end{eqnarray}

\item The class with two Bosonic indices $k$ and $l$ and two Fermionic ones $\alpha$ and $\beta$ contains twelve different forms,
\begin{eqnarray}
\label{con_b2_f2}
&&\epsilon ^{\alpha  \beta } (\eta \sigma \Lambda \Lambda ^k)^l,\qquad 
\epsilon ^{\alpha  \beta } \Lambda ^{k l},\qquad 
\Lambda ^2 \eta ^{k l} \epsilon ^{\alpha  \beta },\qquad
\sigma \Lambda \Lambda  \eta ^{k l} \epsilon ^{\alpha  \beta },\nonumber\\
&&\Lambda ^2 (\sigma ^{k l})^{\alpha  \beta },\qquad 
\sigma \Lambda \Lambda  (\sigma ^{k l})^{\alpha  \beta },\qquad 
\Lambda ^{k o} \eta _{o n} (\sigma ^{n l})^{\alpha  \beta },\nonumber\\
&&\epsilon ^{\alpha  \zeta } \epsilon ^{\beta  \iota } \epsilon ^{k l n o} \eta _{n p} \eta _{o q} \Lambda ^p{}_{\zeta } \Lambda ^q{}_{\iota }, \qquad
\epsilon ^{\alpha  \zeta } \eta _{n o} (\sigma \Lambda ^{o k})^{l \beta } \Lambda ^n{}_{\zeta },\nonumber\\
&&\epsilon ^{\alpha  \zeta } (\eta \sigma \Lambda ^k)^{\beta } \Lambda ^l{}_{\zeta },\qquad 
\epsilon ^{\alpha  \zeta } \epsilon ^{\beta  \iota } \Lambda ^k{}_{\zeta } \Lambda ^l{}_{\iota },\qquad
\eta ^{k l}\epsilon ^{\alpha  \zeta } \eta _{n o} (\eta \sigma \Lambda ^n)^{\beta } \Lambda ^o{}_{\zeta },
\end{eqnarray}
\end{itemize}
where the definitions of abbreviations are given in eq.~(\ref{abbrivation}).

\section{Construction of the effective action of ${\cal S}^{\prime}_{(3)}$}

With the allowed divergent operators and BFNC parameters provided in the above two sections as ingredients, we are now ready to search for the effective action of ${\cal S}^{\prime}_{(3)}$.

For the third class of divergent operators, see eq.~(\ref{part_3}), it is easy to construct $1/2$ supersymmetric invariant forms that are separated into  two categories without and with covariant derivatives. The former contains the five different $1/2$ supersymmetric invariant bases,
\begin{eqnarray}
\label{action_3rd_1}
B_{75}&=&\left(\Lambda ^2 y_{75,1}+\sigma \Lambda \Lambda  z_{75,1}\right) \theta ^4 \Phi ^+,\nonumber\\
B_{76}&=&\left(\Lambda ^2 y_{76,1}+\sigma \Lambda \Lambda  z_{76,1}\right) \theta ^4 \Phi ^+ \Phi ^+,\nonumber\\
B_{77}&=&\left(\Lambda ^2 y_{77,1}+\sigma \Lambda \Lambda  z_{77,1}\right) \theta ^4 \Phi ^+ \Phi ^+ \Phi ^+,\nonumber\\
B_{78}&=&\left(\Lambda ^2 y_{78,1}+\sigma \Lambda \Lambda  z_{78,1}\right) \theta ^4 \Phi ^+ \Phi ^+ \Phi ^+ \Phi ^+,\nonumber\\
B_{79}&=&\left(\Lambda ^2 y_{79,1}+\sigma \Lambda \Lambda  z_{79,1}\right) \theta ^4 \Phi ^+ \Phi ^+ \Phi ^+ \Phi ^+ \Phi ^+,
\end{eqnarray}
and the latter has the six ones as follows,
\begin{eqnarray}
\label{action_3rd_2}
B_{80}&=&\left(\Lambda ^2 y_{80,1}+\sigma \Lambda \Lambda  z_{80,1}\right) \theta ^4 \left(\bar{D}^2 \Phi ^+\right),\nonumber\\
B_{81}&=&\left(\Lambda ^2 y_{81,2}+\sigma \Lambda \Lambda  z_{81,2}\right) \theta ^4 \left(\bar{D}^2 \Phi ^+\right) \Phi ^+\nonumber\\
&&+\left(\Lambda ^2 y_{81,2}+\sigma \Lambda \Lambda  z_{81,2}\right) \epsilon ^{\dot{\alpha } \dot{\beta }} \theta ^4 \left(\bar{D}_{\dot{\alpha }} \Phi ^+\right) \left(\bar{D}_{\dot{\beta }} \Phi ^+\right),\nonumber\\
B_{82}&=&\left(\frac{1}{2} \Lambda ^2 y_{82,2}+\frac{1}{2} \sigma \Lambda \Lambda  z_{82,2}\right) \theta ^4 \left(\bar{D}^2 \Phi ^+\right) \Phi ^+ \Phi ^+\nonumber\\
&&+\left(\Lambda ^2 y_{82,2}+\sigma \Lambda \Lambda  z_{82,2}\right) \epsilon ^{\dot{\alpha } \dot{\beta }} \theta ^4 \left(\bar{D}_{\dot{\alpha }} \Phi ^+\right) \left(\bar{D}_{\dot{\beta }} \Phi ^+\right) \Phi ^+,\nonumber\\
B_{83}&=&\left(\frac{1}{3} \Lambda ^2 y_{83,2}+\frac{1}{3} \sigma \Lambda \Lambda  z_{83,2}\right) \theta ^4 \left(\bar{D}^2 \Phi ^+\right) \Phi ^+ \Phi ^+ \Phi ^+\nonumber\\
&&+\left(\Lambda ^2 y_{83,2}+\sigma \Lambda \Lambda  z_{83,2}\right) \epsilon ^{\dot{\alpha } \dot{\beta }} \theta ^4 \left(\bar{D}_{\dot{\alpha }} \Phi ^+\right) \left(\bar{D}_{\dot{\beta }} \Phi ^+\right) \Phi ^+ \Phi ^+,\nonumber\\
B_{84}&=&\left(\frac{1}{4} \Lambda ^2 y_{84,2}+\frac{1}{4} \sigma \Lambda \Lambda  z_{84,2}\right) \theta ^4 \left(\bar{D}^2 \Phi ^+\right) \Phi ^+ \Phi ^+ \Phi ^+ \Phi ^+\nonumber\\
&&+\left(\Lambda ^2 y_{84,2}+\sigma \Lambda \Lambda  z_{84,2}\right) \epsilon ^{\dot{\alpha } \dot{\beta }} \theta ^4 \left(\bar{D}_{\dot{\alpha }} \Phi ^+\right) \left(\bar{D}_{\dot{\beta }} \Phi ^+\right) \Phi ^+ \Phi ^+ \Phi ^+,\nonumber\\
B_{85}&=&\left(\frac{1}{5} \Lambda ^2 y_{85,2}+\frac{1}{5} \sigma \Lambda \Lambda  z_{85,2}\right) \theta ^4 \left(\bar{D}^2 \Phi ^+\right) \Phi ^+ \Phi ^+ \Phi ^+ \Phi ^+ \Phi ^+\nonumber\\
&&+\left(\Lambda ^2 y_{85,2}+\sigma \Lambda \Lambda  z_{85,2}\right) \epsilon ^{\dot{\alpha } \dot{\beta }} \theta ^4 \left(\bar{D}_{\dot{\alpha }} \Phi ^+\right) \left(\bar{D}_{\dot{\beta }} \Phi ^+\right) \Phi ^+ \Phi ^+ \Phi ^+ \Phi ^+,
\end{eqnarray}
where the parameters $y_{i,j}$ and $z_{i,j}$ are introduced for the definition of $1/2$ supersymmetry invariant bases~\cite{Miao:2013a}.

For the fourth class of divergent operators, we take the  2-point function as an example since it is straightforward to extend the analysis to the other point functions. 
In order to search for the part of the effective action of ${\cal S}^{\prime}_{(3)}$ through analyzing the terms in the fourth class, we list them on the left hand side and their corresponding terms in ${\cal S}^{\prime}_{(3)}$ on the right hand side.

We choose the terms with two $\Phi$'s from eq.~(\ref{part_4_2}),
\begin{eqnarray}
\label{}
D^2 \Phi  \Phi & \Longleftrightarrow & (-\Lambda ^2 y_{1,2}-\sigma \Lambda \Lambda  z_{1,2} ) \epsilon ^{\alpha  \beta } \theta ^4 \left(D_{\alpha } \Phi \right)  (D_{\beta } \Phi  )\nonumber\\
&&+ (\Lambda ^2 y_{1,2}+\sigma \Lambda \Lambda  z_{1,2} ) \theta ^4 \Phi  (D^2 \Phi),\label{ex1_1}\\
D^2 D^2 \Phi  \Phi & \Longleftrightarrow & (\Lambda ^2 y_{38,1}+\sigma \Lambda \Lambda  z_{38,1}) \theta ^4 (D^2 \Phi ) (D^2 \Phi)\label{ex1_2},\\
D^2\partial\partial \Phi  \Phi & \Longleftrightarrow & x_{16,2} (\eta \sigma \Lambda \Lambda ^k)^l \theta ^4 \Phi  \partial _k\partial _l (D^2 \Phi)\nonumber\\
&&+x_{18,2} \Lambda ^{k l} \theta ^4 \Phi  \partial _l\partial _k (D^2 \Phi )\nonumber\\
&&+(\Lambda ^2 y_{22,2}+\sigma \Lambda \Lambda  z_{22,2}) \theta ^4 \Phi  \square (D^2 \Phi )\nonumber\\
&&+x_{16,2} \epsilon ^{\alpha  \beta } (\eta \sigma \Lambda \Lambda ^k)^l \theta ^4 (D_{\beta } \Phi ) \partial _k \partial _l (D_{\alpha } \Phi)\nonumber\\
&&+x_{18,2} \epsilon ^{\alpha  \beta } \Lambda ^{k l} \theta ^4  (D_{\beta } \Phi ) \partial _l\partial _k (D_{\alpha } \Phi )\nonumber\\
&&+(\Lambda ^2 y_{22,2}+\sigma \Lambda \Lambda  z_{22,2}) \epsilon ^{\alpha  \beta } \theta ^4 (D_{\beta } \Phi ) \square (D_{\alpha } \Phi), \label{ex1_3} \\
D^2 D^2 \partial\partial \Phi  \Phi & \Longleftrightarrow & x_{42,1}  (\eta \sigma \Lambda \Lambda ^k )^l \theta ^4  (D^2 \Phi  ) \partial _k\partial _l (D^2 \Phi  )\nonumber\\
&&+x_{44,1} \Lambda ^{k l} \theta ^4  (D^2 \Phi  ) \partial _l\partial _k (D^2 \Phi  )\nonumber\\
&&+ (\Lambda ^2 y_{71,1}+\sigma \Lambda \Lambda  z_{71,1} ) \theta ^4  (D^2 \Phi  ) \square (D^2 \Phi  ).\label{ex1_4}
\end{eqnarray}
The meaning of the above equations is, for instance, that in eq.~(\ref{ex1_1}) the operator $D^2 \Phi  \Phi $ (on the left hand side) in the fourth class corresponds to the terms (on the right hand side) in ${\cal S}^{\prime}_{(3)}$ with two $D$'s and two $\Phi$'s.

Let us interpret the distributions of operators  in eq.~(\ref{ex1_1}) which contains two terms and corresponds to two allowed distributions of covariant derivatives acting on superfields. 
When $D$'s and $\bar{D}$'s act on chiral or antichiral superfields, we should take into account the following identities,
\begin{equation}
\label{chiral_condition}
D_{\alpha } \Phi ^+=0,\qquad \bar{D}_{\dot{\alpha }} \Phi =0.
\end{equation}
Due to eq.~(\ref{chiral_condition}), when $D^2$ acts on two $\Phi$'s, the allowed distributions are as follows,
\begin{equation}
\label{2D_2Phi}
\left(D_{\alpha }\Phi \right)\left(D_{\beta }\Phi \right),\qquad
\left(D^2\Phi \right)\Phi.
\end{equation}

In a 2-point function, 
when $D^2$ acts on a chiral superfield $\Phi$ and an antichiral superfield $\Phi^+$, the non-vanishing contribution emerges from the two $D$'s acting on $\Phi$ but not on $\Phi ^+$, that is, $\left(D^2\Phi \right)\Phi ^+$. 
With this reason and that already demonstrated under eq.~(\ref{condition D}), we can determine all of the allowed distributions of covariant derivatives $D$'s and $\bar{D}$'s acting on superfields for the operators in the fourth class containing 1, 3, 4, 5, and 6 superfields.

In eq.~(\ref{ex1_2})   there is only one allowed distribution when four $D$'s act on two $\Phi$'s, see eq.~(\ref{chiral_condition}).

Let us continue our analysis for  eq.~(\ref{ex1_3}). 
For the Fermionic indices, there are two allowed distributions of two $D$'s acting on two $\Phi$'s as explained in eq.~(\ref{2D_2Phi}).
For the Bosonic indices, we can move all Bosonic derivatives to the front of any of two superfields by using the integration by parts,
\begin{equation}
\label{partial_derivatives}
\int d^4x\left(\partial _k {\bf F}\right){\bf G}=-\int d^4x \,{\bf F}\left(\partial _k {\bf G}\right),
\end{equation}
where ${\bf F}$ and  ${\bf G}$ denote any superfields. 
For three and higher point functions, we can also find the method to arrange the Bosonic derivatives. 
As an example, for the product of superfields like  $\Phi\Phi\Phi^+$,  according to  eq.~(\ref{partial_derivatives}) we can demand  no Bosonic derivatives acting on $\Phi^+$ by moving all the Bosonic derivatives acting on $\Phi^+$ to the places where they only act on the two $\Phi$'s.

In the lines 1, 2, and 3 on the right hand side of eq.~(\ref{ex1_3}), $\Phi  \partial _k\partial _l\left(D^2 \Phi \right)$ 
has only two symmetric Bosonic indices. 
Because the combination of symmetric indices with antisymmetric indices leads to vanishing contributions, we must combine $\Phi  \partial _k\partial _l\left(D^2 \Phi \right)$ with the BFNC parameters that are not antisymmetric with respect to indices $k$ and $l$,
see eq.~(\ref{con_b2}).

In the lines 4, 5, and 6 on the right hand side of eq.~(\ref{ex1_3}), $\left(D_{\beta } \Phi \right) \partial _k\partial _l\left(D_{\alpha } \Phi \right)$ has two Bosonic indices $k$ and $l$ and two Fermionic indices $\alpha$ and  $\beta$. 
Because the Bosonic derivatives have been moved together, the indices $k$ and $l$ are symmetric. 
For the combinations of $\left(D_{\beta } \Phi \right) \partial _k\partial _l\left(D_{\alpha } \Phi \right)$ and the BFNC parameters with two Fermionic indices, let us consider the following term at first,
\begin{equation}
\label{example_2_point_kl_D}
\int d^4x\left(D_{\alpha }\Phi \right)\left(\partial _k\partial _lD_{\beta }\Phi \right).
\end{equation}
By using eq.~(\ref{partial_derivatives}) twice, we can transform eq.~(\ref{example_2_point_kl_D}) into
\begin{equation}
\label{example_2_point_kl_D_2}
-\int d^4x\left(D_{\beta }\Phi \right)\left(\partial _k\partial _lD_{\alpha }\Phi \right),
\end{equation}
where the factor $-1$ appears due to the  exchange of Fermionic numbers. 
Then by using Fierz identity,
\begin{equation}
\label{fierz_identity}
{\bf F}_{\alpha }{\bf G}_{\beta }-{\bf F}_{\beta }{\bf G}_{\alpha }=-\epsilon _{\alpha \beta }\epsilon ^{\gamma \zeta }{\bf F}_{\gamma }{\bf G}_{\zeta },
\end{equation}
we can transform eq.~(\ref{example_2_point_kl_D}) into the following form,
\begin{equation}
\label{example_2_point_kl_D_3}
-\frac{1}{2}\epsilon _{\alpha \beta }\epsilon ^{\gamma \zeta }\int d^4x\left(D_{\gamma }\Phi \right)\left(\partial _k\partial _lD_{\zeta }\Phi \right),
\end{equation}
which is antisymmetric with respect to $\alpha$ and $\beta$. 
As the indices $\alpha$ and $\beta$ of eq.~(\ref{example_2_point_kl_D}) are antisymmetric, so we can only combine them with those BFNC parameters of eq.~(\ref{con_b2_f2}) that are not symmetric with respect to indices $\alpha$ and $\beta$. 
In conclusion, the BFNC parameters that can be combined with eq.~(\ref{example_2_point_kl_D}) should have no antisymmetric indices with respect to $k$ and $l$, and have no symmetric indices with respect to $\alpha$ and $\beta$. 
Moreover, we note that for the following BFNC parameters of eq.~(\ref{con_b2_f2}),
\begin{eqnarray}
\label{example_with_epsilon}
\epsilon ^{\alpha  \zeta } \eta _{n o} (\sigma \Lambda ^{o k})^{l \beta } \Lambda ^n{}_{\zeta },\qquad
\epsilon ^{\alpha  \zeta }(\eta \sigma \Lambda ^k)^{\beta } \Lambda ^l{}_{\zeta },\qquad
\eta ^{k l}\epsilon ^{\alpha  \zeta } \eta _{n o} (\eta \sigma \Lambda ^n)^{\beta } \Lambda ^o{}_{\zeta },
\end{eqnarray}
we can transform them to $(\eta \sigma \Lambda ^k)^l$ or $\sigma \Lambda \Lambda  \eta ^{k l}$ when we combine them with antisymmetric indices of $\epsilon _{\alpha \beta }$.  
Therefore, we conclude that all possible terms constructed by $D^2 \partial\partial \Phi  \Phi $ in the effective action of ${\cal S}^{\prime}_{(3)}$  are covered by the right hand side of eq.~(\ref{ex1_3}).

Let us consider now the three lines on the right hand side of  eq.~(\ref{ex1_4}). 
As there is only one allowed form for the distribution of four $D$'s acting on two $\Phi$'s, which is similar to eq.~(\ref{ex1_2}), we need only to consider the possible combinations of the BFNC parameters with two Bosonic derivatives. 
As a result,  eq.~(\ref{ex1_4}) can be dealt with by the same way as that to the lines 1, 2, and 3 of eq.~(\ref{ex1_3}).

Next we list the remaining terms of 2-point functions in the fourth class. 
For the terms with one $\Phi$ and one $\Phi ^+$, the correspondence is as follows, 
\begin{eqnarray}
\label{}
\partial\partial \Phi  \Phi ^+  & \Longleftrightarrow & \left(16 \Lambda ^2 y_{23,3}+16 \sigma \Lambda \Lambda  z_{23,3}\right) \theta ^4 \Phi  \square \Phi ^+\label{ex2_1}, \\
\partial D  \bar{D}\Phi  \Phi ^+  & \Longleftrightarrow & \left(-8 i \Lambda ^2 y_{23,3}-8 i \sigma \Lambda \Lambda  z_{23,3}\right) (\bar{\sigma }^k)^{\dot{\alpha } \beta } \theta ^4 \left(D_{\beta } \Phi \right) \partial _k\left(\bar{D}_{\dot{\alpha }} \Phi ^+\right)\label{ex2_2}, \\
D^2 \overline{D}^2 \Phi  \Phi ^+  & \Longleftrightarrow & \left(\Lambda ^2 y_{23,3}+\sigma \Lambda \Lambda  z_{23,3}\right) \theta ^4 \left(D^2 \Phi \right) \left(\bar{D}^2 \Phi ^+\right)\label{ex2_3}, \\
D^2 \Phi  \Phi ^+  & \Longleftrightarrow & \left(\Lambda ^2 y_{49,1}+\sigma \Lambda \Lambda  z_{49,1}\right) \theta ^4 \left(D^2 \Phi \right) \Phi ^+\label{ex2_4}, \\
D^2 \partial\partial \Phi  \Phi ^+  & \Longleftrightarrow & x_{55,1} (\eta \sigma \Lambda \Lambda ^k)^l \theta ^4 \left(D^2 \Phi \right) \partial _k\partial _l\Phi ^+\nonumber\\
&&+x_{56,1} \Lambda ^{k l} \theta ^4 \left(D^2 \Phi \right) \partial _k\partial _l\Phi ^+\nonumber\\
&&+\left(\Lambda ^2 y_{73,1}+\sigma \Lambda \Lambda  z_{73,1}\right) \theta ^4 \left(D^2 \Phi \right) \square \Phi ^+\label{ex2_5}, \\
D^2 \partial\partial\partial\partial \Phi  \Phi ^+  & \Longleftrightarrow & x_{59,1} (\eta \sigma \Lambda \Lambda ^k)^l \theta ^4 \left(D^2 \Phi \right) \square \partial _k\partial _l\Phi ^+\nonumber\\
&&+x_{60,1} \Lambda ^{k l} \theta ^4 \left(D^2 \Phi \right) \square \partial _k\partial _l\Phi ^+\nonumber\\
&&+\left(\Lambda ^2 y_{67,1}+\sigma \Lambda \Lambda  z_{67,1}\right) \theta ^4 \left(D^2 \Phi \right) \square \square \Phi ^+\label{ex2_6},
\end{eqnarray}

For the operator $\partial\partial \Phi\Phi^+$, we find from eq.~(\ref{ex2_1}) that there are only two BFNC parameters $\Lambda ^2 \eta ^{k l}$ and $\sigma \Lambda \Lambda  \eta ^{k l}$ that can be combined with $\partial _k\partial _l$.  
Moreover, from eq.~(\ref{con_b2}) we know that there are two other BFNC parameters $(\eta \sigma \Lambda \Lambda ^k)^l$ and $\Lambda ^{k l}$ that also have indices $k$ and $l$.

Similarly, the indices of the operator $\partial D  \bar{D}\Phi  \Phi ^+$ (see eq.~(\ref{ex2_2})) are $k$, $\dot{\alpha}$, and ${\beta}$, according to eq.~(\ref{con_b1_f2}) we can also combine this operator with the following BFNC parameters,
\begin{eqnarray}
\label{newBFNCparameters}
\Lambda ^{k l} \eta _{l n} (\bar{\sigma }^n )^{\dot{\alpha } \beta },\qquad 
\eta _{n l}  (\bar{\sigma }^l )^{\dot{\alpha } \beta }  (\eta \sigma \Lambda \Lambda ^n )^k,\qquad
\eta _{n l}  (\bar{\sigma }^l )^{\dot{\alpha } \beta }  (\eta \sigma \Lambda \Lambda ^k )^n.
\end{eqnarray}
However,  such combinations (that $\partial D  \bar{D}\Phi  \Phi ^+$ combines with the above three BFNC parameters) do not exist in the action ${\cal S}^{\prime}_{(3)}$. 
Let us give an explanation. 
By examining the terms in eqs.~(\ref{ex2_1}), (\ref{ex2_2}), and (\ref{ex2_3}), we observe that they form a $1/2$ supersymmetry invariant subset, see ref.~\cite{Miao:2013a} for the details. 
This indicates that the $1/2$ supersymmetry invariance plays a fundamental role for constructing the effective action of ${\cal S}^{\prime}_{(3)}$. 
To explore this, we combine all of the allowed BFNC parameters with operators $\partial\partial \Phi  \Phi ^+$, $\partial D \overline{D} \Phi  \Phi ^+$, and $D^2 \overline{D}^2 \Phi  \Phi ^+$ that appear in eqs.~(\ref{ex2_1}), (\ref{ex2_2}), and (\ref{ex2_3}), respectively. 
As their indices should be matched up, not all of BFNC parameters are involved. Then we use all of the possible combinations to construct $1/2$ supersymmetry invariant subsets in light of the method introduced in ref.~\cite{Miao:2013a}. 
We discover that the $1/2$ supersymmetry invariance is not maintained if the new combinations of the BFNC parameters (see eq.~(\ref{newBFNCparameters})) and the operator $\partial D \overline{D} \Phi  \Phi ^+$ are added, which gives the reason that ${\cal S}^{\prime}_{(3)}$ does not contain these new combinations. 
In other words, not only the indices should be matched up, but also the  $1/2$ supersymmetry invariance should be satisfied.

The combinations in eqs.~(\ref{ex2_4}), (\ref{ex2_5}), and (\ref{ex2_6}) are easy to be understood in terms of the above statement on the combinations of BNFC parameters and operators with two $\Phi$'s.

We list the last two terms of 2-point functions with two $\Phi^+$'s (see eq.~(\ref{part_4_2})) and their corresponding terms in ${\cal S}^{\prime}_{(3)}$,
\begin{eqnarray}
\label{}
\partial\partial \Phi ^+ \Phi ^+  & \Longleftrightarrow & \left(\Lambda ^2 y_{39,1}+\sigma \Lambda \Lambda  z_{39,1}\right) \theta ^4 \Phi ^+ \square \Phi ^+\label{ex3_1}\\
\partial\partial\partial\partial \Phi ^+ \Phi ^+  & \Longleftrightarrow & x_{43,1}  (\eta \sigma \Lambda \Lambda ^k )^l \,\theta ^4 \Phi ^+ \square \partial _k\partial _l\Phi ^+,\nonumber\\
&&+x_{45,1} \Lambda ^{k l} \theta ^4 \Phi ^+ \square \partial _k\partial _l\Phi ^+\nonumber\\
&&+\left(\Lambda ^2 y_{72,1}+\sigma \Lambda \Lambda  z_{72,1}\right) \theta ^4 \Phi ^+ \square \square \Phi ^+\label{ex3_2}.
\end{eqnarray}
From eq.~(\ref{ex3_2}) we see that if there exists an operator composed of  the product of 4 Bosonic derivatives, $\partial_k$, $\partial_l$, $\partial_n$, and $\partial_o$, the BFNC parameters that can be combined with it should not be antisymmetric with respect to any two indices of $k$, $l$, $n$, and $o$.

By using eqs.~(\ref{con_b0})-(\ref{con_b2_f2}) we can determine the part of the effective action of ${\cal S}^{\prime}_{(3)}$ that corresponds to the 2-point operators. 
In conclusion, we point out that the following forms which can be constructed by using the operators in the fourth class are not contained in ${\cal S}^{\prime}_{(3)}$, 
\begin{eqnarray}
\label{2_point_not_in_1_loop}
\Lambda ^{k l}\theta ^4 \Phi ^+ \partial _k\partial _l\Phi ^+, \qquad
 (\eta \sigma \Lambda \Lambda ^k )^l  \theta ^4 \Phi ^+ \partial _k\partial _l\Phi ^+.
\end{eqnarray}

With the same consideration as above for 1-, 3-, 4-, 5-, and 6-point functions, i.e., the divergent operators in the fourth class (see eqs.~(\ref{part_4_1}) and (\ref{part_4_3})-(\ref{part_4_6})), we can construct all allowed parts of the effective action of ${\cal S}^{\prime}_{(3)}$ for the operators in the fourth class 
and find that most of them are contained in ${\cal S}^{\prime}_{(3)}$, except the following two forms, 
\begin{eqnarray}
\label{3_point_not_in_1_loop}
\Lambda ^{k l}\theta ^4 \Phi ^+\Phi ^+ \partial _k\partial _l\Phi ^+, \qquad
 (\eta \sigma \Lambda \Lambda ^k )^l  \theta ^4 \Phi ^+ \Phi ^+\partial _k\partial _l\Phi ^+.
\end{eqnarray}

It is easy to construct $1/2$ supersymmetric invariant forms by using eqs.~(\ref{2_point_not_in_1_loop}) and (\ref{3_point_not_in_1_loop}),
\begin{eqnarray}
\label{action_4th}
B_{86}&=&x_{86,1} \Lambda ^{k l} \theta ^4 \Phi ^+ \partial _k\partial _l\Phi ^+\nonumber\\
B_{87}&=&x_{87,1} \left(\eta \sigma \Lambda \Lambda ^k\right)^l \theta ^4 \Phi ^+ \partial _k\partial _l\Phi ^+\nonumber\\
B_{88}&=&x_{88,1} \Lambda ^{k l} \theta ^4 \Phi ^+ \Phi ^+ \partial _k\partial _l\Phi ^+\nonumber\\
B_{89}&=&x_{89,1} \left(\eta \sigma \Lambda \Lambda ^k\right)^l \theta ^4 \Phi ^+ \Phi ^+ \partial _k\partial _l\Phi ^+
\end{eqnarray}
where the parameter $x_{i,j}$ is introduced for the definition of $1/2$ supersymmetry invariant bases~\cite{Miao:2013a}.

In a word, we have found all possible terms in the effective action of ${\cal S}^{\prime}_{(3)}$ by using the most general divergent operators and BFNC parameters.
In this process, we have taken into account the requirement of $1/2$ supersymmetry. 
We thus obtain the whole of the effective action systematically by classifying the operators according to the number of superfields and by comparing them with ${\cal S}^{\prime}_{(3)}$.
Our conclusion is that under the second order approximation of BFNC parameters the effective action of ${\cal S}^{\prime}_{(3)}$ is
\begin{equation}
\label{gen_EA}
\int d^8z\left(\sum _{i=1}^{74} c_i B_i\right)+\int d^8z\left(\sum _{i=75}^{89} c_i B_i\right),
\end{equation}
where the coefficients $c_i$'s, $i=75, \cdots, 89$, are introduced to make 
$\int d^8z\left(\sum_{i=75}^{89}c_i B_i\right)$ 
to have $U(1)_{\rm R}$ $R$-symmetry and $U(1)_{\Phi}$ flavor symmetry and are listed in eq.~(\ref{coefficients_c}).
Note that the first part of the above equation covered by ${\cal S}^{\prime}_{(3)}$ (see eq.~(\ref{action_1})) holds a large part of the terms in the effective action.
We are now ready to give a BFNC Wess-Zumino model which is renormalizable at all loops in perturbation theory.

\section{Conclusion and Outlook}

By comparing  the part of the effective action constructed from the third and fourth classes with the terms in ${\cal S}^{\prime}_{(3)}$, we find that almost all terms in this part of the  effective action are contained in ${\cal S}^{\prime}_{(3)}$, but only the terms in eqs.~(\ref{action_3rd_1}), (\ref{action_3rd_2}) and (\ref{action_4th}) are not contained. 
We now deduce that ${\cal S}^{\prime}_{(3)}$ plus these terms (eqs.~(\ref{action_3rd_1}), (\ref{action_3rd_2}) and (\ref{action_4th})) is renormalizable at all loops in perturbation theory,
\begin{equation}
\label{finalaction}
{\cal S}^{\prime}_{(3)}+\int d^8z\left(\sum_{i=75}^{89}c_i B_i\right).
\end{equation}
Alternatively, eq.~(\ref{finalaction}) can be rewritten as follows in terms of eq.~(\ref{action_1}),
\begin{equation}
\label{alternativeaction}
{\cal S}^{\prime}_{\rm WZ}+\int d^8z\left(\sum _{i=1}^{89} c_i B_i\right).
\end{equation}

Finally, we make a comment that our above investigation is limited to the second order of BFNC parameters. 
For deriving a renormalizable action at a higher order of BFNC parameters, we construct all divergent operators in terms of the symmetry analysis, and then deduce all allowed terms of an effective action using the divergent operators. At last we find out $1/2$ supersymmetry invariant subsets and their corresponding bases. The result obtained in this way must be renormalizable at all loops and at a higher order of BFNC parameters in perturbation theory.

\section*{Acknowledgments}
Y-GM would like to thank
H.P. Nilles of the University of Bonn for kind hospitality.
This work was supported in part by the Alexander von Humboldt Foundation under a renewed research program, by the National Natural
Science Foundation of China under grant No.11175090 and
by the Ministry of Education of China under grant No.20120031110027. At last, the authors would like to thank the anonymous referee for the helpful comments that indeed improve this work greatly.

\newpage
\section*{Appendix}

\appendix

\setcounter{equation}{0}
\renewcommand\theequation{A\arabic{equation}}

The coefficients $c_i$'s, where $i=1, \cdots, 89$, are listed below.
\begin{align}
\label{coefficients_c}
c_1&=g^2 m \left(m^*\right)^3,&c_2&=g^3 \left(m^*\right)^3,&c_3&=g m \left(m^*\right)^2,&c_4&=g^2 \left(m^*\right)^2,\nonumber\\
c_5&=m m^*,&c_6&=g m^*,&c_7&=m g^*,&c_8&=1,\nonumber\\
c_9&=g m m^*,&c_{10}&=g^2 m^*,&c_{11}&=m,&c_{12}&=g,\nonumber\\
c_{13}&=g \left(m^*\right)^2,&c_{14}&=m^*,&c_{15}&=g^*,&c_{16}&=g^4 \left(m^*\right)^5,\nonumber\\
c_{17}&=g m^*,&c_{18}&=g^2 \left(m^*\right)^2,&c_{19}&=g^2 \left(m^*\right)^3,&c_{20}&=1,\nonumber\\
c_{21}&=1,&c_{22}&=g^2 \left(m^*\right)^2,&c_{23}&=g^2 \left(m^*\right)^3,&c_{24}&=g \left(m^*\right)^2,\nonumber\\
c_{25}&=m^*,&c_{26}&=g^*,&c_{27}&=g,&c_{28}&=g^3 \left(m^*\right)^3,\nonumber\\
c_{29}&=g,&c_{30}&=g m^2 \left(m^*\right)^3,&c_{31}&=g \left(m^*\right)^3,&c_{32}&=g,\nonumber\\
c_{33}&=g \left(m^*\right)^3,&c_{34}&=m \left(m^*\right)^3,&c_{35}&=m \left(m^*\right)^3,&c_{36}&=g,\nonumber\\
c_{37}&=g,&c_{38}&=m^2 m^*,&c_{39}&=m \left(m^*\right)^2,&c_{40}&=m^2 g^*,\nonumber\\
c_{41}&=m g^* m^*,&c_{42}&=g^2 m \left(m^*\right)^3,&c_{43}&=g^2 \left(m^*\right)^4,&c_{44}&=m,\nonumber\\
c_{45}&=m^*,&c_{46}&=g^*,&c_{47}&=m \left(g^*\right)^2,&c_{48}&=m \left(g^*\right)^2,\nonumber\\
c_{49}&=m^2 \left(m^*\right)^2,&c_{50}&=m^2 g^* m^*,&c_{51}&=m^2 \left(g^*\right)^2,&c_{52}&=g \left(m^*\right)^3,\nonumber\\
c_{53}&=g m \left(m^*\right)^3,&c_{54}&=m g^*,&c_{55}&=g^2 m \left(m^*\right)^4,&c_{56}&=m m^*,\nonumber\\
c_{57}&=g m \left(m^*\right)^3,&c_{58}&=m g^*,&c_{59}&=g^2 \left(m^*\right)^3,&c_{60}&=1,\nonumber\\
c_{61}&=g \left(m^*\right)^3,&c_{62}&=g^*,&c_{63}&=g^*,&c_{64}&=g^*,\nonumber\\
c_{65}&=g^*,&c_{66}&=m \left(g^*\right)^2,&c_{67}&=1,&c_{68}&=g^*,\nonumber\\
c_{69}&=m \left(g^*\right)^2,&c_{70}&=m g^*,&c_{71}&=m,&c_{72}&=m^*,\nonumber\\
c_{73}&=m m^*,&c_{74}&=m g^*,&c_{75}&=g m^2 \left(m^*\right)^4,&c_{76}&=m^2 \left(m^*\right)^3,\nonumber\\
c_{77}&=m^2 g^* \left(m^*\right)^2,&c_{78}&=m^2 \left(g^*\right)^2 m^*,&c_{79}&=m^2 \left(g^*\right)^3,&c_{80}&=g^3 \left(m^*\right)^5,\nonumber\\
c_{81}&=g^2 \left(m^*\right)^4,&c_{82}&=g \left(m^*\right)^3,&c_{83}&=\left(m^*\right)^2,&c_{84}&=g^* m^*,\nonumber\\
c_{85}&=\left(g^*\right)^2,&c_{86}&=m \left(m^*\right)^2,&c_{87}&=g^2 m \left(m^*\right)^5,&c_{88}&=m g^* m^*,\nonumber\\
c_{89}&=g m \left(m^*\right)^4.
\end{align}

\end{document}